%% file: main.tex
\documentclass[sigplan,screen]{acmart}
\usepackage[utf8]{inputenc}
\usepackage{microtype}
\copyrightyear{2027}
\acmYear{2027}
\setcopyright{cc}
\setcctype{by}
\acmConference[CPP '27]{Proceedings of the 16th ACM SIGPLAN International Conference on Certified Programs and Proofs}{January 11--12, 2027}{Mexico City, Mexico}
\acmBooktitle{Proceedings of the 16th ACM SIGPLAN International Conference on Certified Programs and Proofs (CPP '27), January 11--12, 2027, Mexico City, Mexico}
\acmDOI{10.1145/3871849.3871884}
\acmISBN{979-8-4007-3146-4/2027/01}
\acmSubmissionID{poplws27cppmain-p73-p}
\received{2026-09-10}
\received[accepted]{2026-11-10}

\input{macros}

\makeatletter
\@ifundefined{KV@Gin@artifact}{\define@key{Gin}{artifact}[]{}}{}
\makeatother

\begin{document}
\title{Machine-Checked Grammar-Constrained Decoding}

\author{Manu Bhat}
\orcid{0009-0003-8783-7793}
\authornote{These authors contributed equally to this work.}
\affiliation{%
  \institution{University of California San Diego}
  \city{La Jolla}
  \state{California}
  \country{USA}}
\email{mbhat@ucsd.edu}
\correspondingauthor

\author{Arnav Dandu}
\orcid{0009-0006-7580-2608}
\authornotemark[1]
\affiliation{%
  \institution{University of California San Diego}
  \city{La Jolla}
  \state{California}
  \country{USA}}
\email{adandu@ucsd.edu}

\author{Kanghee Park}
\orcid{0009-0005-7983-233X}
\affiliation{%
  \institution{University of California San Diego}
  \city{La Jolla}
  \state{California}
  \country{USA}}
\email{kap022@ucsd.edu}

\author{Loris D'Antoni}
\orcid{0000-0001-9625-4037}
\affiliation{%
  \institution{University of California San Diego}
  \city{La Jolla}
  \state{California}
  \country{USA}}
\email{ldantoni@ucsd.edu}
\renewcommand{\shortauthors}{Bhat, Dandu, Park, D'Antoni}

\begin{abstract}
\input{sections/abstract}
\end{abstract}



\maketitle

\input{sections/introduction}
\input{sections/overview}
\input{sections/formal_model}
\input{sections/lexing}
\input{sections/algorithm}
\input{sections/correctness}
\input{sections/discovered_issues}
\input{sections/related_work}
\input{sections/limitations_conclusion}

\input{sections/data_availability}

\appendix
\input{sections/appendix_lean}
\input{sections/appendix_theorem_map}

\begin{acks}
The project is supported in part by a Microsoft Faculty Fellowship; a UCSD JSOE Scholarship; Schmidt foundation; and NSF under grants CCF-2422214, CCF-2506134, CCF-2446711. Any opinions, findings, and conclusions or recommendations expressed in this publication are those of the authors, and do not necessarily reflect the views of the sponsoring entities. Loris D’Antoni holds concurrent appointments as a Professor at the University of California San Diego and as a Code Metal Scholar. This paper describes work performed at the University of California San Diego and is not associated with Code Metal. We thank Keith Johnson for his helpful discussions and feedback on the paper.
\end{acks}

{\scriptsize
\bibliographystyle{ACM-Reference-Format}
\bibliography{references}
}

\end{document}

%% file: macros.tex
\usepackage{amsmath}
\usepackage{booktabs}
\usepackage{listings}
\usepackage{tabularx}
\usepackage{xspace}
\usepackage{mathtools}
\usepackage{microtype}
\usepackage{enumitem}
\usepackage{newunicodechar}
\usepackage{upgreek}
\usepackage{tikz}
\usetikzlibrary{automata,positioning,arrows.meta}
\usepackage{algorithm}
\usepackage{algpseudocode}
\usepackage{stfloats}
\usepackage{xurl} 

\usepackage{xcolor}

\definecolor{leanbg}{HTML}{FAFAFC}
\definecolor{leanframe}{HTML}{D7DAE0}
\definecolor{leankw}{HTML}{7C3AED}
\definecolor{leanstr}{HTML}{0F766E}
\definecolor{leancom}{HTML}{3F6212}
\definecolor{leannum}{HTML}{B45309}

\newunicodechar{α}{\ensuremath{\upalpha}}
\newunicodechar{β}{\ensuremath{\upbeta}}
\newunicodechar{γ}{\ensuremath{\upgamma}}
\newunicodechar{π}{\ensuremath{\uppi}}
\newunicodechar{σ}{\ensuremath{\upsigma}}
\newunicodechar{→}{\ensuremath{\rightarrow}}
\newunicodechar{∀}{\ensuremath{\forall}}
\newunicodechar{∃}{\ensuremath{\exists}}
\newunicodechar{∈}{\ensuremath{\in}}
\newunicodechar{∉}{\ensuremath{\notin}}
\newunicodechar{×}{\ensuremath{\times}}
\newunicodechar{≠}{\ensuremath{\neq}}
\newunicodechar{∧}{\ensuremath{\land}}
\newunicodechar{↔}{\ensuremath{\leftrightarrow}}
\newunicodechar{Γ}{\ensuremath{\Gamma}}
\newunicodechar{¬}{\ensuremath{\neg}}
\newunicodechar{∨}{\ensuremath{\lor}}
\newunicodechar{∅}{\ensuremath{\emptyset}}

\newcommand{\lean}[1]{\ifmmode\text{\normalfont\ttfamily\detokenize{#1}}\else{\normalfont\ttfamily\detokenize{#1}}\fi}
\makeatletter
\newcommand{\leanid}[1]{{\normalfont\ttfamily\expandafter\leanid@scan\detokenize{#1}\leanid@end}}
\def\leanid@end{\leanid@end}
\def\leanid@scan#1{%
  \ifx#1\leanid@end
  \else
    \ifnum`#1>64 \ifnum`#1<91 \allowbreak\fi\fi
    #1%
    \ifnum`#1=95 \allowbreak\fi
    \expandafter\leanid@scan
  \fi}
\makeatother

\newcommand{\GCDname}{Grammar-constrained decoding\xspace}
\newcommand{\greatgramma}{\textsc{GreatGramma}\xspace}

\newcommand{\lang}{\mathcal{L}}
\newcommand{\term}[1]{\ensuremath{\textsc{#1}}}

\newcommand{\eos}{\mathsf{eos}}

\theoremstyle{plain}
\newtheorem{theorem}{Theorem}

\newtheorem{informaltheorem}[theorem]{Informal theorem}

\theoremstyle{definition}

\lstdefinelanguage{Lean}{
  sensitive=true,
  morekeywords={
    def, theorem, lemma, abbrev, example,
    inductive, structure, class, instance,
    namespace, section, variable,
    where, deriving,
    match, with,
    if, then, else,
    let, in, fun,
    forall, exists,
    by, simp, rw, exact,
    constructor, cases,
    have, show,
    open, import,
    mutual, termination_by,
    Type, Prop
  },
  morecomment=[l]{--},
  morecomment=[s]{/-}{-/},
  morestring=[b]",
  literate=
    {→}{{$\rightarrow$}}1
    {←}{{$\leftarrow$}}1
    {↔}{{$\leftrightarrow$}}1
    {⇒}{{$\Rightarrow$}}1
    {∀}{{$\forall$}}1
    {∃}{{$\exists$}}1
    {∈}{{$\in$}}1
    {∉}{{$\notin$}}1
    {⊆}{{$\subseteq$}}1
    {∪}{{$\cup$}}1
    {∩}{{$\cap$}}1
    {×}{{$\times$}}1
    {⊢}{{$\vdash$}}1
    {⊥}{{$\bot$}}1
    {⊤}{{$\top$}}1
    {≤}{{$\le$}}1
    {≥}{{$\ge$}}1
    {≠}{{$\neq$}}1
    {¬}{{$\neg$}}1
    {∧}{{$\land$}}1
    {∨}{{$\lor$}}1
    {λ}{{$\lambda$}}1
    {α}{{$\alpha$}}1
    {β}{{$\beta$}}1
    {γ}{{$\gamma$}}1
    {σ}{{$\sigma$}}1
    {Γ}{{$\Gamma$}}1
    {∅}{{$\emptyset$}}1
    {α}{{$\upalpha$}}1 {β}{{$\upbeta$}}1 {γ}{{$\upgamma$}}1 {π}{{$\uppi$}}1 {σ}{{$\upsigma$}}1 {→}{{$\rightarrow$}}1 {∀}{{$\forall$}}1 {∃}{{$\exists$}}1 {∈}{{$\in$}}1 {∉}{{$\notin$}}1 {∪}{{$\cup$}}1 {×}{{$\times$}}1 {≠}{{$\neq$}}1 {∧}{{$\land$}}1 {↔}{{$\leftrightarrow$}}1 {Γ}{{$\Gamma$}}1 {¬}{{$\neg$}}1 {∨}{{$\lor$}}1,
}

\setlist[itemize]{leftmargin=1.6em,itemsep=0.25ex,topsep=0.6ex}
\setlist[enumerate]{leftmargin=1.8em,itemsep=0.25ex,topsep=0.6ex}

%% file: sections/abstract.tex
Grammar-constrained decoding (GCD) restricts a language model's next-token choices so that its output remains extendable to a string accepted by a user-provided grammar. \greatgramma is a compositional GCD algorithm built from automata-theoretic components. We formalize its executable checker in Lean 4 and prove it sound and complete under explicit assumptions on the vocabulary, lexer, parser, and whitespace. To our knowledge, this is the first machine-checked end-to-end correctness proof for a grammar-constrained decoding algorithm for LLMs.

The formalization finds an end-of-stream bug in the published lexer construction and makes explicit a universal-assumption needed for the algorithm. The development also includes reusable verified components for finite-state automata and transducers, pushdown automata, parser overapproximations, and graph search.

%% file: sections/introduction.tex
\section{Introduction}
\label{sec:introduction}

Large language models (LLMs) are increasingly used to generate structured artifacts rather than free-form text: JSON objects, SQL queries, API calls, configuration files, proof scripts, and program fragments~\cite{chen2021codex,gao2024texttosql,schick2023toolformer,patil2024gorilla,first2023baldur}. In these settings, even a small syntactic mistake can make the output unusable. A missing quote, an extra comma, or a malformed keyword may be enough for a downstream parser or tool to reject the result. Prompting can reduce these errors, but it cannot guarantee that they never happen. Figure~\ref{fig:gcd-motivation} shows an LLM producing malformed JSON despite understanding the requested structure.

\GCDname (GCD) addresses this problem by forcing the LLM to generate only outputs that belong to a user-provided grammar. At each decoding step, GCD considers the current token prefix and uses a \emph{checker} to mask every token that cannot be extended to a valid sequence in the grammar. The LLM then samples only among tokens that can still lead to a string accepted by the grammar.

\begin{figure}[t]
  \centering
  \scriptsize
  \setlength{\tabcolsep}{2pt}
  \begin{tabularx}{\linewidth}{@{}lX@{}}
    \toprule
    Prompt & Return a JSON object with city, days, and temperature unit. \\
    \midrule
    Grammar &
    $\begin{array}[t]{rcl}
      \mathit{obj} &::=& \texttt{\{}\,\texttt{"city"}\,\texttt{:}\,\mathit{str}\,\texttt{,}\,\texttt{"days"}\,\texttt{:}\,\mathit{num}\,\texttt{,}\,\texttt{"unit"}\,\texttt{:}\,\mathit{unit}\,\texttt{\}}\\
      \mathit{unit} &::=& \texttt{"C"}\mid\texttt{"F"}
    \end{array}$ \\
    \midrule
    Unconstrained &
    \begin{tabular}[t]{@{}l@{}}\texttt{\{city:"Paris", days:3, unit:C,\}}\\ rejected by the grammar.\end{tabular} \\
    With GCD &
    \begin{tabular}[t]{@{}l@{}}\texttt{\{"city":"Paris","days":3,"unit":"C"\}}
    \end{tabular} \\
    \bottomrule
  \end{tabularx}
  \caption{A grammar-constrained decoding scenario.  The mask keeps generation on prefixes that can still be completed to the requested JSON shape.}
  \Description{A table contrasts an unconstrained malformed JSON response with a grammar-constrained response that has valid quotes and punctuation.}
  \label{fig:gcd-motivation}
\end{figure}

At first glance, building such a checker may look like a straightforward application of parsing. In practice, two technical issues make an efficient construction considerably more subtle.

\smallskip
\noindent\textbf{Token--terminal misalignment.}
LLMs generate \emph{tokens}, which are variable-length sequences of characters determined by the tokenizer. Grammars instead operate over characters that are grouped by a lexer into \emph{terminals}, which are then consumed by the parser. Token boundaries and terminal boundaries therefore need not line up. A single token may contain an entire terminal, such as \texttt{true}; only part of one, such as \texttt{tru}; or several terminals at once, such as \texttt{\}]}.

\smallskip
\noindent\textbf{Decoding efficiency.}
A GCD checker runs at every decoding step, where latency matters. Modern LLM vocabularies contain tens of thousands of tokens, so simply trying every candidate token by running the lexer and parser from scratch is far too expensive (taking seconds instead of milliseconds). Practical GCD algorithms therefore perform substantial preprocessing and exploit the structure of the vocabulary, lexer, and parser so that the online next-token check is cheap.

These optimizations are exactly where things become difficult. Several systems have made GCD practical for code, structured data, and tool-call generation~\cite{geng2023gcd,poesia2022synchromesh,willard2023efficient,beurerkellner2024domino,ugare2024syncode,dong2024xgrammar}. However, \citet{park2025flexible} showed that a number of optimized GCD algorithms are not correct: some produce sequences that do not belong to the language of the grammar, others mask tokens that should remain available, while still others employ preprocessing algorithms that do not terminate even on simple grammars.

To avoid these ad-hoc optimizations, \citet{park2025flexible} proposed \greatgramma, an efficient GCD algorithm built by composing standard automata-theoretic constructions. During preprocessing, \greatgramma computes how each LLM token interacts with partially read lexemes and records which terminals the token may expose. During decoding, it combines this information with the current parser state to determine which tokens still admit a grammatical completion. 

Despite its simple design, \greatgramma achieves performance comparable to other GCD implementations while often reducing preprocessing time. Its modular structure also makes it a natural target for formal verification: rather than verify a monolithic optimized implementation, we formalize and compose the automata constructions from which its checker is built.

Our formalization largely validates the modular structure proposed by \citet{park2025flexible}, but it also shows why a machine-checked proof is useful. Specifically, we discovered a bug in the published lexing construction that can cause the end-of-stream token to be accepted while the lexer is still in the middle of an unfinished terminal. We isolate the source of the bug, give a corrected construction, and verify the corrected version in Lean~4.

Our main result is an end-to-end machine-checked correctness theorem for an executable \greatgramma checker. Under explicit assumptions on the vocabulary, lexer, parser, and treatment of whitespace, we prove that the checker is both \emph{sound} and \emph{complete}: after any token prefix, it allows exactly those next tokens that can still be extended to a grammatical output. To the best of our knowledge, this is the first machine-checked end-to-end correctness proof for a grammar-constrained decoding algorithm for LLMs.

\paragraph{Contributions.}
Concretely, the paper makes the following contributions.
\begin{itemize}
\item \textbf{A machine-checked end-to-end correctness theorem for \greatgramma} (Sections~\ref{sec:overview} and \ref{sec:correctness}). We formalize its core constructions and prove the executable checker sound and complete for the next-token decision under explicit assumptions. The Lean development is available at \url{https://github.com/large-loris-models/great-gramma-formalization}.
\item \textbf{A bug, a fix, and explicit assumptions} (Section~\ref{sec:findings}). We diagnose and fix the published lexer's end-of-stream bug, and make explicit the implicit conditions the proof requires, including the universal-separator assumption on whitespace.
\item \textbf{Reusable Lean infrastructure} (Sections~\ref{sec:formal-model}--\ref{sec:algorithm}). Our development includes partial finite-state automata and transducers, nondeterministic pushdown automata, stack-free parser overapproximations, token--character--terminal interfaces, and verified finite graph search.
\end{itemize}

Sections~\ref{sec:related-work}, \ref{sec:limitations}, and~\ref{sec:conclusion} cover related work, formalization lessons and limitations, and conclusions.

%% file: sections/overview.tex
\section{GCD and \greatgramma}
\label{sec:overview}

This section introduces the decoding problem and the checker specification, then
gives an intuitive account of \greatgramma's finite summary of lexer behavior.
The formal machine models and assumptions follow in Section~\ref{sec:formal-model}.

\subsection{Models, Grammars, and Constrained Generation}
\label{subsec:gcd-interface}

A language model generates one token at a time from a finite vocabulary $V$ and a distinguished end-of-stream symbol $\eos$.  Given a prefix $w = v_1\cdots v_n$, it assigns each next symbol a probability
\begin{equation}
  \Pr(v \mid w) \qquad\text{for } v \in V \cup \{\eos\}.
  \label{eq:llm-next-token}
\end{equation}
Sampling appends the chosen token, or stops at $\eos$; the next distribution is conditioned on the extended prefix.

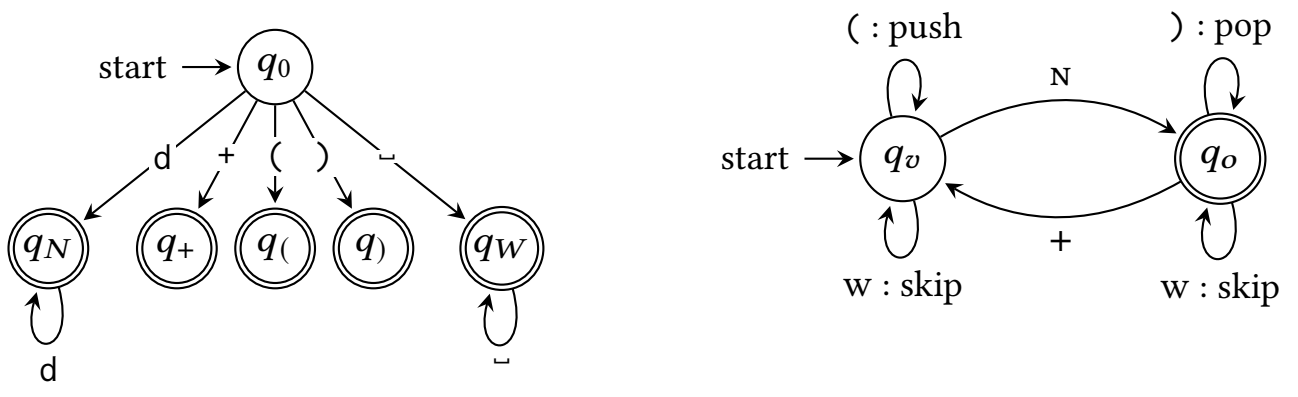
\begin{figure*}[!t]
  \newcommand{\toprowheight}{2.95cm}
  \footnotesize
  \medskip
  \begin{minipage}[c][\toprowheight][c]{0.14\textwidth}
    \begin{center}
    \begin{tabular}{@{}ll@{}}
      \toprule
      Terminal & Regex \\
      \midrule
      $\term{n}$ & \texttt{[0-9]+} \\
      $\mathord{+}$ & \texttt{+} \\
      $\mathord{(}$ & \texttt{(} \\
      $\mathord{)}$ & \texttt{)} \\
      $\term{w}$ & \texttt{\textvisiblespace+} \\
      \bottomrule
    \end{tabular}
    \smallskip

    (a) Terminals
    \end{center}
  \end{minipage}\hfill
  \begin{minipage}[c][\toprowheight][c]{0.12\textwidth}
    \begin{center}
    \begin{tabular}{@{}l@{}}
      \toprule
      \texttt{1}, \texttt{2}, \texttt{12} \\
      \texttt{+}, \texttt{(}, \texttt{)} \\
      \texttt{\textvisiblespace}, \texttt{1+}, \texttt{)+} \\
      \bottomrule
    \end{tabular}
    \smallskip

    (b) Vocabulary
    \end{center}
  \end{minipage}\hfill
  \begin{minipage}[c][\toprowheight][c]{0.17\textwidth}
    \begin{center}
    $\begin{array}{r@{\;}c@{\;}l}
      E &::=& A \mid A \mathbin{\texttt{+}} E \\[2pt]
      A &::=& \term{n} \mid \texttt{(}\,E\,\texttt{)} \\
    \end{array}$
    \smallskip

    (c) Grammar ($\term{w}$ ignored)
    \end{center}
  \end{minipage}\hfill
  \begin{minipage}[c][\toprowheight][c]{0.28\textwidth}
    \begin{center}
    \begin{tikzpicture}[
        >=stealth, shorten >=1pt,
        every node/.style={font=\scriptsize},
        every state/.style={minimum size=14pt,inner sep=1pt},
        lbl/.style={fill=white,inner sep=0.7pt,font=\scriptsize}]
      \node[state,initial left] (q0) at (1.7,1.2) {$q_0$};
      \node[state,accepting] (qN) at (0.2,0.0) {$q_N$};
      \node[state,accepting] (qP) at (1.05,0.0) {$q_+$};
      \node[state,accepting] (qL) at (1.7,0.0) {$q_($};
      \node[state,accepting] (qR) at (2.35,0.0) {$q_)$};
      \node[state,accepting] (qW) at (3.2,0.0) {$q_W$};
      \path[->]
        (q0) edge node[lbl,pos=0.5] {\texttt{d}} (qN)
        (q0) edge node[lbl,pos=0.5] {\texttt{+}} (qP)
        (q0) edge node[lbl,pos=0.5] {\texttt{(}} (qL)
        (q0) edge node[lbl,pos=0.5] {\texttt{)}} (qR)
        (q0) edge node[lbl,pos=0.5] {\texttt{\textvisiblespace}} (qW)
        (qN) edge[loop below] node {\texttt{d}} (qN)
        (qW) edge[loop below] node {\texttt{\textvisiblespace}} (qW);
    \end{tikzpicture}
    \smallskip

    (d) Lexer automaton (\texttt{d}=digit)
    \end{center}
  \end{minipage}\hfill
  \begin{minipage}[c][\toprowheight][c]{0.22\textwidth}
    \begin{center}
    \begin{tikzpicture}[
        >=stealth, shorten >=1pt, node distance=2.1cm,
        on grid, auto,
        every node/.style={font=\scriptsize},
        every state/.style={minimum size=16pt,inner sep=1pt}]
      \node[state,initial] (val) {$q_v$};
      \node[state,accepting,right=of val] (op) {$q_o$};
      \path[->]
        (val) edge[bend left] node[above] {$\term{n}$} (op)
        (op) edge[bend left] node[below] {$\mathord{+}$} (val)
        (val) edge[loop above] node {$\texttt{(}$ : push} (val)
        (op) edge[loop above] node {$\texttt{)}$ : pop} (op)
        (val) edge[loop below] node {$\term{w}$ : skip} (val)
        (op) edge[loop below] node {$\term{w}$ : skip} (op);
    \end{tikzpicture}
    \smallskip

    (e) Parser PDA ($q_v$ wants a value, $q_o$ has one;
    $\term{w}$ is a no-op)
    \end{center}
  \end{minipage}
  \caption{A \greatgramma example for a small grammar of parenthesized sums of
  numbers (e.g.\ \texttt{1}, \texttt{1+2}, \texttt{(1+2)+3}).  Terminal regexes (a)
  define a lexer automaton (d); the vocabulary (b) is analyzed against it to build
  a table of realizable sequence heads; and the parser (e), a pushdown
  automaton because of the parentheses, decides which heads still lead to an
  accepted continuation.  Whitespace $\term{w}$ is a universal separator: for the parser it
  is a no-op (a self-loop at every state, (e)), so \texttt{1+2} and \texttt{1 + 2}
  parse identically and no valid string's meaning depends on it.}
  \label{fig:greatgramma-example}
\end{figure*}

Tokens map to strings over the character alphabet $\Sigma$.  We write
$\mathrm{detok}:V^*\to\Sigma^*$ for the map extended by concatenation, so distinct
token sequences can represent the same character string. For instance,
$\mathrm{detok}([\text{\texttt{1}},\text{\texttt{2}}])$ and
$\mathrm{detok}([\text{\texttt{12}}])$ both yield \texttt{12} in
Figure~\ref{fig:greatgramma-example}.

A grammar $G$ describes a language $\lang_G \subseteq \Sigma^*$, a set of character strings.  In practice, its rules are built from \emph{terminals} rather than individual characters.  Each terminal is a lexical class defined by a regular expression over characters.
Given such a description, recognizing a string proceeds in two stages: lexing and parsing.
First, a lexer reads the character stream and, whenever it recognizes a maximal matching substring (a \emph{lexeme}), emits the corresponding terminal.  This longest-match rule is \emph{maximal munch}, and deciding where one lexeme ends takes a single character of lookahead, a point we return to in Section~\ref{sec:lexing}.  Second, a pushdown automaton (the parser) decides whether the resulting terminal sequence belongs to the grammar.  Throughout, ``terminal'' refers to the symbol the parser reads and ``lexeme'' to the concrete characters that produced it.

To summarize, the target semantics is
\[
  \text{tokens}
  \xrightarrow{\mathrm{detok}}
  \text{chars}
  \xrightarrow{\text{lexer}}
  \text{terminals}
  \xrightarrow{\text{parser}}
  \text{accept}.
\]
A complete token sequence is valid exactly when this detokenization, lexing, and parsing pipeline succeeds.

Figure~\ref{fig:greatgramma-example} illustrates all components.  For the token
sequence $w=[\text{\texttt{1}},\text{\texttt{+}},\text{\texttt{2}}]$ we have
$\mathrm{detok}(w)=\text{\texttt{1+2}}$, which the lexer tokenizes as $\term{n}\,{+}\,\term{n}$;
the parser accepts $\term{n}\,{+}\,\term{n}$ as a well-formed sum.

\GCDname forces an LLM to produce outputs that adhere to a given grammar.
It works on top of an unmodified model, acting only at the sampling step.
Specifically, when the LLM provides the probability distribution $\Pr(v \mid w)$ for a prefix $w$, GCD should mask exactly the next tokens that cannot lead to a complete output in the grammar.
Masking sets the probability of those tokens to $0$ and renormalizes the distribution over the remaining tokens.
For a grammar $G$ with language $\lang_G$, we call a token prefix \emph{viable} when it has some valid continuation:
\begin{equation}
  \mathsf{viable}_G(w) \quad\text{if}\quad \exists u\in V^*.\; \mathrm{detok}(wu)\in\lang_G.
  \label{eq:viable-prefix}
\end{equation}
This predicate is the ground-truth mask that a checker must efficiently compute.

We model a checker as a function $C_G : V^* \times (V \cup \{\eos\}) \to \mathsf{Bool}$, where $C_G(w,v)=\mathsf{true}$ means that $v$ is allowed after prefix $w$, and is masked otherwise.  For an ordinary token $v\in V$, a correct checker allows $v$ after $w$ exactly when $\mathsf{viable}_G(wv)$ holds.  We say a checker is \emph{sound} if every token it allows from a viable prefix keeps the prefix viable, and \emph{complete} if every viable token is allowed.

Our goal and ultimate result is to prove that the verified \greatgramma checker implements this specification.  We state the assumptions precisely in Section~\ref{sec:formal-model}.
\begin{informaltheorem}[End-to-end correctness of \greatgramma]
Assume the vocabulary, lexer, parser, and whitespace satisfy the conditions described in this paper. For every viable token prefix $w$ and every candidate token $v\in V$, the verified \greatgramma checker allows $v$ exactly when there exists some sequence of future LLM tokens $u\in V^*$ such that
\[
\mathrm{detok}(wvu)\in\lang_G.
\]
Moreover, the checker allows $\eos$ after $w$ exactly when
\[
\mathrm{detok}(w)\in\lang_G.
\]
\end{informaltheorem}
In other words, \greatgramma computes the correct next-token mask induced by the grammar.
\subsection{\greatgramma at a High Level}
\label{subsec:greatgramma-high-level}

In Section~\ref{sec:algorithm} we describe the algorithm in detail; here we give a brief overview.

Recall that a grammar is built from a lexer and a parser where the lexer consumes a character sequence and produces a terminal sequence that the parser accepts or rejects. The main challenge a GCD checker faces, then, is characterizing the terminal sequences the lexer can produce at any given time and the subset of those sequences the parser accepts.  We call a terminal sequence that the lexer can produce from a given state a \emph{realizable terminal sequence}.  Ultimately, every token that does not start a realizable terminal sequence compatible with the current parser state must be masked.

To make realizable terminal sequences more concrete, we illustrate a simple grammar in Figure~\ref{fig:greatgramma-example}.  Suppose the current token prefix is $w=[\text{\texttt{1}}]$.  Its detokenized text leaves the lexer in state $q_N$ (the digit is an unfinished number), while the parser is still in its start state $q_v$ with an empty stack, since no terminal has been emitted yet.  Consider the candidate token \texttt{+}.  Reading \texttt{+} ends the pending number lexeme, so the lexer emits the terminal $\term{n}$ and starts an operator lexeme.  Hence, every realizable terminal sequence it can still lead to begins with $\term{n}\,{+}$.  At least one of those sequences is accepted, namely $\term{n}\,{+}\,\term{n}$ (the lexing of the complete string \texttt{1+2}), so \texttt{+} keeps a path to a grammatical output open and should be allowed.  The candidate \texttt{)} instead leads only to sequences beginning $\term{n}\,)$, and from an empty stack the parser cannot read \texttt{)} at all, since it has no matching \texttt{(} to pop.  No completion is accepted, so \texttt{)} should be masked.  The very same \texttt{)} \emph{is} allowed if we change the prefix to $[\text{\texttt{(}},\text{\texttt{1}}]$: now the stack holds an open parenthesis, the \texttt{)} closes it, and a string such as \texttt{(1)} is accepted.

The key observation in \greatgramma is that the set of realizable terminal sequences can be summarized by looking ahead only a finite amount.  We call a \emph{realizable sequence head} the terminals a token immediately produces, followed by one further terminal that the lexer can produce next; such a next terminal is \emph{single-producible}.  For instance, if the current lexer state is $q_N$ and the candidate token is \texttt{+}, the token immediately produces $\term{n}$ (committing the pending number) and leaves the lexer poised to produce a \texttt{+} next, so it exposes the head $\term{n}\,{+}$.  A token can also produce nothing yet: from $q_N$ the token \texttt{1} only extends the number, so its head is just $\term{n}$.

We show that, under explicit assumptions, the lexer can produce a terminal sequence if and only if it begins with one of these realizable sequence heads, modulo whitespace that the parser ignores.  Hence a token has an accepted continuation exactly when at least one of its associated heads is allowed by the parser.  \greatgramma uses this fact in two phases.  In an offline preprocessing phase, it builds a token-spanner table recording which heads each token can expose from each lexer state.  During online decoding, the checker tests those heads against the current parser state and then uses the inverse token-spanner table to translate the surviving heads back into the tokens that should be allowed.

The checker combines three parts: token-level lexing (Sections~\ref{sec:formal-model}
and~\ref{sec:lexing}); the inverse token-spanner table (Section~\ref{sec:algorithm});
and parser preprocessing with online masking (Section~\ref{sec:algorithm}). This
decomposition supports a proof from localized lemmas.

%% file: sections/formal_model.tex
\section{Formal Models}
\label{sec:formal-model}
From here on, we use Lean's type notation: \(α\) and \(β\) correspond to the
character alphabet \(\Sigma\) and token vocabulary \(V\), while \(γ\), \(π\),
and \(σ\) denote terminal, stack-symbol, and state types.

\subsection{Automata, Transducers, and Pushdown Automata}

A grammar is realized by two machines.  A lexer turns characters into terminals,
and a parser decides whether a terminal sequence belongs to the grammar.  We
model the lexer with deterministic finite-state machines and the parser with a
\emph{nondeterministic} pushdown automaton.

\begin{lstlisting}
structure FSA (α σ) where
  start : σ
  step : σ → α → Option σ
  accept : Finset σ

structure FST (α γ σ) where
  start : σ
  step : σ → α → Option (σ × List γ)
  accept : Finset σ

structure PDA (γ π σ) where
  start : σ
  step : σ → γ → Finset (List π × List π × σ)
  accept : Finset σ
\end{lstlisting}

An \lean{FSA} is a deterministic acceptor over an input alphabet \(α\).  Its
transition function is \emph{partial}: a missing transition returns \lean{none},
and the evaluator propagates that failure.  We use a partial function rather than
Mathlib's total \lean{DFA} because the checker must tell a still-viable input
apart from one that has already gone wrong.
An \lean{FST} is the analogous deterministic transducer.  It fails via
\lean{none}, and on success returns a next state together with a finite list of
output symbols.

A \lean{PDA} models the parser.  Its transition relation is nondeterministic and
stack-sensitive: on a terminal \(T\), a move matches a prefix of the stack,
replaces it with another stack fragment, and jumps to a new control state.  Its
evaluator maps a finite set of configurations to a finite set of configurations,
and failure leaves that set empty.

Three notions recur throughout.  A machine \emph{allows} an input when running on
it has not failed, and \emph{accepts} an input when it additionally ends in an
accepting state.  A pruned FSA has every state reachable from its start and able
to reach acceptance.  A pruned PDA has an accepting continuation from every
reachable configuration.  For a pruned machine the allowed inputs are exactly the
prefixes of accepted inputs,
\[
  \mathsf{allows}(w) \iff \exists u.\ \mathsf{accepts}(w \mathbin{+\!\!+} u),
\]
an equivalence (\lean{FSA.pruned_prefixLanguage}) we use in the lexer and
soundness proofs.  In the parser of Figure~\ref{fig:greatgramma-example}, a single
number \(\term{n}\) is both allowed and accepted; \(\term{n}\,{+}\) is allowed but not accepted,
since the parser then expects another operand; and \({+}\,\term{n}\) is not even allowed,
since the start state has no transition on \({+}\).

\subsection{Vocabularies and Detokenization}

The vocabulary is the interface between the model's tokens and the characters a
grammar refers to.

\begin{lstlisting}
class Vocabulary (α : Type) (β : Type) where
  flatten : β → List α
  embed   : α → β
  fe      : ∀ a, flatten (embed a) = [a]
  empty   : ∀ b, flatten b ≠ []
\end{lstlisting}

Here \lean{flatten} detokenizes a token into its characters, and the \lean{empty}
law forbids tokens that detokenize to nothing.  The \lean{embed} field with the
law \lean{flatten (embed a) = [a]} records \greatgramma's \emph{singleton-token}
assumption that every single character is itself a token.

\subsection{Checkers}

The online object the paper verifies is a next-token \lean{Checker}, a function
\(\mathtt{List}\ β \to \mathtt{Ch}\ β \to \mathtt{Bool}\) that reads the token prefix generated so far and answers
whether a candidate next symbol is locally permitted.  The candidate has the
\(\eos\)-extended token type \lean{Ch}~\(β\), which adds a distinguished
end-of-stream symbol \(\eos\) to the ordinary tokens, so one interface decides both
ordinary tokens and the choice to stop generating.

\subsection{Lexer Specifications}

The checker takes a \emph{lexer specification}, which pairs a character automaton
with a terminal label on each accepting state, together with a few sanity conditions
relating the two.

\begin{lstlisting}
structure LexerSpec (α γ σ) where
  automaton : FSA α σ
  term      : σ → Option γ
  hterm     : ∀ s, s ∈ automaton.accept ↔ (term s).isSome
  term_inj  : ...
  term_surj : ...
\end{lstlisting}

From such a specification, \lean{BuildLexingFST} produces the lexing transducer over
\(\eos\)-extended characters and terminals, and the checker runs its composition
\lean{BuildDetokLexer} with detokenization directly on tokens
(Section~\ref{sec:lexing}).
Here \lean{automaton} recognizes the union of the terminal regex languages;
\lean{term} labels states, and \lean{hterm} says exactly the accepting states carry
labels (and every label is attached to an accepting state).

\subsection{Assumptions: The Universal Separator}
\label{subsec:assumptions}

The end-to-end theorem holds under a small bundle of assumptions.  The lexer recognizes no empty
lexeme, and the lexer and parser are pruned.  The key assumption is the
\emph{universal separator}.

It has two halves, organized around a distinguished whitespace character and its
terminal.  On the lexer side, a whitespace character belongs to no lexeme other than
the whitespace terminal, so it always ends whatever lexeme precedes it.  On the
parser side, the whitespace terminal is a no-op:
\begin{lstlisting}
def ParserIgnoresTerminal (P : PDA γ π σ) (white : γ) : Prop :=
  ∀ q, P.step q white = {([], [], q)}
\end{lstlisting}
So the checker may place a separator between terminals freely, without changing what
the parser does.

The separator makes the algorithm's preprocessing possible.  At a high level, the lexer's possible future terminal sequences form an infinite set, and because a separator can always be slipped in between lexemes, that set collapses to a finite summary (Section~\ref{subsec:realizable-sequences}).  For instance, a JSON-like instance uses newline as its separator.  This condition already applies to the original paper's construction, and making it an explicit hypothesis is part of our contribution.  More generally, applicability depends on the grammar's lexical rules, and whitespace-sensitive languages such as Python and YAML may not satisfy the assumption (Section~\ref{sec:limitations}).

The hypotheses used by the final theorem are bundled into one structure (the
singleton-token assumption is the exception, since it already lives in the
\lean{Vocabulary} class):
\begin{lstlisting}
structure GCDAssumptions
    (spec : LexerSpec α γ σa) (P : PDA γ π σp)
    (tnonwhite twhite : α)
    (qnonwhite qwhite : σa) where
  hempty        : [] ∉ spec.automaton.accepts
  lexer_pruned  : spec.automaton.pruned
  parser_pruned : P.pruned
  whitespace    : GCDWhitespaceAssumption
    spec P tnonwhite twhite qnonwhite qwhite
\end{lstlisting}
Each later correctness statement is parameterized by a value of this type, so the
theorems make their reliance on these assumptions explicit.

%% file: sections/lexing.tex
\section{Verified Lexing}
\label{sec:lexing}

The checker's first job is to turn the tokens the model produces into the
terminal stream the parser consumes.  We build this lexing layer in three steps:
a formal specification of what the lexer should do, an executable transducer
that the checker runs, and a proof that the two agree.  We then derive the
property the rest of the checker depends on, a finite description of which
terminal sequences the lexer can still produce from a given state.

\subsection{One-Lookahead Lexer Semantics}

Lexing here follows \emph{maximal munch} with a single character of lookahead: the
lexer extends the current lexeme as far as it can and commits a terminal only when the
next character cannot continue it. These constraints allow
the lexer to compile to a simple finite-state transducer, which is the executable form the checker
runs.

We use the lexer specification of Section~\ref{sec:formal-model}, a character
automaton \(\mathcal{A}\) whose accepting states are labeled by the terminals they
recognize.  Write \(\mathcal{L}(\mathcal{A})\) for the lexemes \(\mathcal{A}\)
accepts and \(\mathcal{L}_{\mathrm{prefix}}(\mathcal{A})\) for their prefixes, and
for a lexeme \(u\in\mathcal{L}(\mathcal{A})\) let \(\tau(u)\in\gamma\) be its
terminal label.

Following \citet{park2025flexible}, the lexer is a partial function on
EOS-extended character strings,
\(\mathsf{Lex}\colon \mathsf{Ch}(\alpha)^* \rightharpoonup
\mathsf{Ch}(\gamma)^*\times\alpha^*\), sending a consumed string to the
terminals committed so far and the residual buffer of the current, unfinished
lexeme.  It is defined by recursion, starting from
\(\mathsf{Lex}(\epsilon)=(\epsilon,\epsilon)\).  Given
\(\mathsf{Lex}(w)=(T,\ w_r)\) and a next symbol \(c\), a character or
\(\eos\), and writing \(w_r c\) for \(w_r\mathbin{+\!\!+}[c]\), the value of
\(\mathsf{Lex}(wc)\) is the first case below that applies, and undefined if none
does:
\begin{enumerate}[label=(\arabic*),leftmargin=1.9em,itemsep=0.2ex]
  \item \emph{extend}: \((T,\ w_r c)\) if
        \(w_r c\in\mathcal{L}_{\mathrm{prefix}}(\mathcal{A})\); the buffer can still
        grow, so maximal munch keeps reading.
  \item \emph{commit}: \((T\,\tau(w_r),\ [c])\) if
        \(w_r\in\mathcal{L}(\mathcal{A})\) and \([c]\in\mathcal{L}_{\mathrm{prefix}}(\mathcal{A})\);
        \(c\) cannot extend the lexeme but can start the next one, so \(w_r\) is
        committed as \(\tau(w_r)\) and \(c\) opens a fresh buffer.
  \item \emph{flush}: \((T\,\tau(w_r)\,\eos,\ \epsilon)\) if \(c=\eos\)
        and \(w_r\in\mathcal{L}(\mathcal{A})\); the stream ends on a complete lexeme,
        committed before \(\eos\).
  \item \emph{stop}: \((T\,\eos,\ \epsilon)\) if \(c=\eos\) and
        \(w_r=\epsilon\); the stream ends with an empty buffer.
\end{enumerate}
Once the lexer rejects, the rejection propagates through the rest of \(w\).  Lean
formalizes \(\mathsf{Lex}\) as the inductive relation \lean{PartialLexRel}, with one
constructor per case plus a base case for the empty run:
\begin{lstlisting}
inductive PartialLexRel (spec) :
    chars -> terms -> unlexed -> Prop
  | nil                -- empty run
  | step_char_continue -- extend  (1)
  | step_char_commit   -- commit  (2)
  | step_eos           -- flush   (3)
  | step_nil_eos       -- stop    (4)
\end{lstlisting}
Each constructor carries the side conditions of its case; the full statements are in
Appendix~\ref{app:lean-lexing}.

\subsection{The Executable Lexer and Its Correctness}

The relation \lean{PartialLexRel} formalizes the lexer's behavior but is not executable.  The
checker instead runs \lean{BuildLexingFST}, which compiles the lexer specification into
a transducer over \(\eos\)-extended characters and terminals (Algorithm~2 of Park et
al.~\cite{park2025flexible}).  Its step function consumes the current lexer state
and one extended character, then either rejects or returns the next state and any
terminals emitted by that step.  Here \lean{A} is \lean{spec.automaton}, and
\lean{s} is the automaton state inside
\lean{q} (the full definition is in Appendix~\ref{app:lean-lexing}):
\newpage
\begin{lstlisting}
def step (q : LexingState σ) (c : Ch α) :=
  let s := LexingState.src spec q
  match c with
  | .char c =>
    -- c extends the lexeme; emit nothing
    if h : (A.step s c).isSome then
      some (.id ((A.step s c).get h), [])
    -- c ends a lexeme; commit and restart
    else if h : s ∈ A.accept then
      let T := (spec.term s).get ((spec.hterm s).mp h)
      (A.step A.start c).map fun q' =>
        (.id q', [.char T])
    else none
  | .eos =>
    -- flush a pending lexeme, then stop
    if h : s ∈ A.accept then
      let T := (spec.term s).get ((spec.hterm s).mp h)
      some (start, [.char T, .eos])
    -- otherwise stop at the start state
    else if q = start then some (start, [.eos])
    else none
\end{lstlisting}

Correctness of this layer says the transducer and the specification describe the
same lexer.

\begin{informaltheorem}[Lexing equivalence]
\label{thm:lexing-equiv}
Suppose the lexer spec recognizes no empty lexeme and is pruned.  Then on
every character string, the executable transducer \lean{BuildLexingFST} and the
one-lookahead relation \lean{PartialLexRel} commit the same sequence of
terminals and leave the same residual buffer unlexed.
\end{informaltheorem}

The two directions are \lean{PartialLexRel_to_LexingFST} and
\lean{LexingFST_to_PartialLexRel} (Appendix~\ref{app:lean-lexing}).  Prunedness is what makes the
executable transducer tractable. In particular, the extend case must decide whether the grown
buffer \(w_r c\) can still reach a legal lexeme, and if every reachable automaton
state can reach an accepting one, that far-off question collapses to the local check
``does \(\mathcal{A}\) have a transition on \(c\) at all.''  From there, matching the
two transducers is a long but routine induction on the cases.

\paragraph{A bug exposed by the equivalence proof.}
In attempting to prove the equivalence, we discovered a bug in the published lexing construction.  Specifically, the original lexer did not have a distinguished start state. As a result, the lexer illegally allowed \(\eos\) whenever the lexing automaton ever returned to the start state. The full bug, counterexample, and fix are discussed in Section~\ref{sec:findings}. This perfectly highlights the ability of formalization to discover intricate edge cases that testing alone may skim over.

\subsection{Detokenization and Composition}

The lexer consumes characters, but the checker is driven by tokens, so we
put a detokenizer in front of it.  Detokenization is a one-state transducer that
flattens each token to its characters:
\begin{lstlisting}
def BuildDetokenizingFST : FST β α Unit :=
  { start := (),
    step := fun _ tok => some ((), Vocabulary.flatten tok),
    accept := \{()\} }
\end{lstlisting}
We remark that our detokenizer deviates slightly from \greatgramma, as our detokenizer is based on a deterministic FST but \greatgramma utilizes a nondeterministic FST construction with $\epsilon$ transitions. However, we argue the difference is ultimately unimportant.  Detokenization is a well-known construction, and each token flattens to exactly one string, so the deterministic form computes the same map.  We use it purely for convenience and to avoid introducing nondeterministic FSTs.  The property the rest of the development uses is that the composed machine \lean{BuildDetokLexer} is equivalent to flattening the
token prefix and then running the lexer (\lean{detokenizer_comp}), so facts about the character lexer lift to the composed one.

\subsection{Realizable Tails}
\label{subsec:realizable-sequences}

For a candidate token, the checker must know which terminal sequences the lexer can
still produce after processing the token from current state \(q\).  However, the set of realizable terminal sequences is
typically infinite, so the checker cannot work with it directly.

The universal separator is
what makes a finite, computable summary possible. A whitespace separator can always be slipped in
to close off one lexeme and start the next, so once the lexer commits a first terminal
from \(q\), further terminals can be appended one at a time by interspersing whitespace.  The only real constraint
is therefore \emph{which terminal the lexer can commit first} from \(q\); everything
after it is free, up to whitespace the parser ignores. The intuition is that the first terminal is genuinely constrained by the current lexer state because the unlexed buffer is only the prefix of some subset of terminals, but from there we can use whitespace to reset the lexer state and allow for any next terminal.

For example, from state \(q_N\) in Figure~\ref{fig:greatgramma-example}, with a
partial number buffered, the only terminal the lexer can commit next is \(\term{n}\).
Every continuation from \(q_N\) thus begins with \(\term{n}\), never \texttt{+} or
\texttt{(}.  However, once that first terminal is pinned down, the rest of the sequence
need not be enumerated.

We make this precise.  Let \(\mathsf{terms}_q(c)\) be the terminals \(\mathsf{Lex}\)
commits when run from state \(q\) on character string \(c\) (if it is not rejected),
and define
\[
  \mathsf{realizable}(q) = \{\, \mathsf{terms}_q(c) \mid c \in \mathsf{Ch}(\alpha)^* \,\},
\]
\[
  \mathsf{singleProducible}(q) = \{\, t \mid \exists c.\ \mathsf{terms}_q(c) = [\,t\,] \,\},
\]
the \emph{realizable terminal sequences} and \emph{single-producible terminals} from \(q\)
(\(\mathrm{Re}\) and \(\mathrm{Prod}(q)\) in \citet{park2025flexible}).  As the
intuition above anticipates, once whitespace is disregarded a realizable terminal
sequence is determined by its first terminal.  The theorem below makes this precise.

\begin{informaltheorem}[Realizable tails]
\label{thm:realizable-tails}
Under the lexing and whitespace assumptions, a terminal sequence is realizable
from a lexer state, modulo whitespace, exactly when it is empty or its first
terminal is single-producible from that state
(\leanid{tailModdedRealizableSequences_eq_singleProducibleHead}).
\end{informaltheorem}

The forward direction follows from the singleton-token assumption. Any nonempty
realizable terminal sequence must begin with a terminal producible on its own.

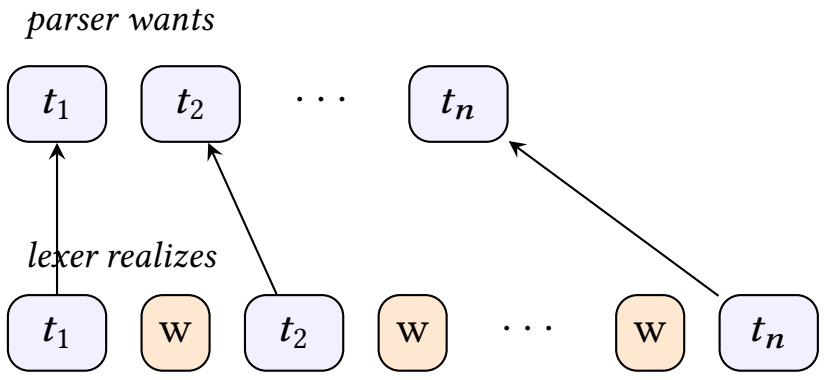
\begin{figure}[!h]
  \centering
  \begin{tikzpicture}[
      >=stealth, font=\footnotesize, node distance=2.2mm,
      term/.style={draw, rounded corners, minimum height=5mm, minimum width=6.5mm, fill=blue!6, inner sep=1pt},
      sep/.style={draw, rounded corners, minimum height=5mm, minimum width=4.5mm, fill=orange!18, inner sep=1pt},
      lab/.style={font=\scriptsize\itshape}]
    \node[term] (a1) {$t_1$};
    \node[term, right=of a1] (a2) {$t_2$};
    \node[right=of a2] (ad) {$\cdots$};
    \node[term, right=of ad] (an) {$t_n$};
    \node[term, below=10mm of a1] (b1) {$t_1$};
    \node[sep, right=of b1] (w1) {$\term{w}$};
    \node[term, right=of w1] (b2) {$t_2$};
    \node[sep, right=of b2] (w2) {$\term{w}$};
    \node[right=of w2] (bd) {$\cdots$};
    \node[sep, right=of bd] (wn) {$\term{w}$};
    \node[term, right=of wn] (bn) {$t_n$};
    \draw[->] (b1) -- (a1);
    \draw[->] (b2) -- (a2);
    \draw[->] (bn) -- (an);
    \node[lab, anchor=south west] at ([yshift=0.5mm]a1.north west) {parser wants};
    \node[lab, anchor=south west] at ([yshift=0.5mm]b1.north west) {lexer realizes};
  \end{tikzpicture}
  \caption{Interspersing the universal separator.  Splicing the whitespace terminal
  \(\term{w}\) (orange) between targets commits each preceding lexeme and resets the
  lexer, so adjacent terminals cannot merge; the parser ignores \(\term{w}\) and
  recovers the original sequence (arrows).}
  \label{fig:interspersing}
\end{figure}

The converse is the substantive direction.  Given a target terminal sequence
\(T=t_1\cdots t_n\) whose first terminal is single-producible from the current state,
we must exhibit an actual character string the lexer turns into that sequence.  The
construction splices the whitespace terminal between successive targets,
\(t_1\,\term{w}\,t_2\,\term{w}\cdots \term{w}\,t_n\)
(Figure~\ref{fig:interspersing}).  Each \(\term{w}\) commits the preceding lexeme
under maximal munch and returns the lexer to the clean post-separator state
\(q_{\term{w}}\), and the parser ignores the inserted separators.

Our goal is to construct a token sequence whose lexer output is
\(T\) with whitespace inserted between its terminals.  We start with a token
sequence witness that produces \(t_1\), since \(t_1\) is single-producible.  We then
massage this witness into the desired one using a carefully chosen sequence of
exchange lemmas that build on one another.
\begin{itemize}
  \item \lean{exchangeNonWhitespace}: a non-whitespace terminal that is
        single-producible from the current state can be produced by a run ending in
        the clean state \(q_{\term{w}}\).
  \item \lean{exchangeEOS}: likewise \(\eos\), when single-producible, can be
        produced by a run ending in \(q_{\term{w}}\).
  \item \lean{exchangeWhitespace}: from \(q_{\term{w}}\), the separator
        \(\term{w}\) can be emitted while reaching any state other than the start
        and whitespace states, steering the lexer to the next target terminal.
  \item \lean{exchangeWhitespaceEOS}: the separator followed by \(\eos\) can be
        produced when the run should end.
\end{itemize}
Together these let the construction emit one target terminal, reset to
\(q_{\term{w}}\), steer across a separator to where the next target begins, and
repeat.  They provide the glue needed to concatenate the witness for one terminal
with the witness for the next.  An induction on \(n\) assembles the
full run and proves the converse.

This is the main lexer-side result used by the soundness proof of
Section~\ref{sec:correctness}.  We note that
this is where we found formal verification to be the most helpful since, while the
construction is intuitive at a high level, its many endpoint cases are difficult
to check reliably by hand.

%% file: sections/algorithm.tex
\section{The \greatgramma Algorithm}
\label{sec:algorithm}

With lexing in place, we can describe the checker.  For a candidate token it must
answer one question: after taking this token, can the lexer go on to emit some
terminal sequence that the current parser accepts?  There are infinitely many
such futures and re-running a parser on each is far too slow, so \greatgramma
precomputes the lexer behavior and the stack-independent parser behavior into
finite tables and leaves only stack-sensitive work online.  This section
describes those tables and the online check that uses them.

\subsection{Realizable Sequence Heads}

The key to a finite summary is Theorem~\ref{thm:realizable-tails}: modulo
whitespace, a terminal sequence is realizable as soon as its first terminal is
single-producible.  To capture everything a candidate token can do to the
terminal stream, we record the terminals it emits immediately followed by one
further producible terminal.  After that, arbitrary tails are realizable and need
not be enumerated.

We call such a finite prefix a \emph{realizable sequence
head}\footnote{The \greatgramma paper calls these objects ``realizable
sequences''.  We rename them to keep the finite prefix distinct from the full,
possibly unbounded sequence it stands for.}: a sequence
\(T \mathbin{+\!\!+} [t]\) arising from some token at some state, where \(T\) is
the (possibly empty) terminal sequence the token emits and \(t\) is a
single-producible terminal afterwards.
Instantiated at the current state, this is the forward token-spanner table of
Figure~\ref{fig:token-spanner}, computed for the running example of
Figure~\ref{fig:greatgramma-example}.  From lexer state \(q_N\) the token
\texttt{+} commits the pending number and emits \(\term{n}\), then leaves the lexer poised
to produce \texttt{+} next, contributing the head \(\term{n}\,{+}\).  The token \texttt{1}
instead extends the current number without committing and emits nothing, and since
\(\term{n}\) is the only terminal producible next from \(q_N\), it contributes the single
head \(\term{n}\).  The differing head lengths are the row-\(q_N\) entries of the table.

\begin{figure}[t]
  \centering
  \scriptsize
  \setlength{\tabcolsep}{3pt}
  \begin{tabular}{@{}c*{9}{c}@{}}
    \toprule
    Lexer state / token & \texttt{1} & \texttt{2} & \texttt{12} & \texttt{+} & \texttt{(} & \texttt{)} & \texttt{\textvisiblespace} & \texttt{1+} & \texttt{)+} \\
    \midrule
    $q_0$   & $\term{n}$  & $\term{n}$  & $\term{n}$  & $+$   & $($   & $)$   & $\term{w}$ & $\term{n}\,{+}$ & $)\,{+}$ \\
    $q_N$   & $\term{n}$  & $\term{n}$  & $\term{n}$  & $\term{n}\,{+}$ & $\term{n}\,($ & $\textbf{\textsc{n}}\,\mathbf{)}$ & $\term{n}\,\term{w}$ & $\term{n}\,{+}$ & $\term{n}\,)\,{+}$ \\
    $q_+$   & ${+}\,\term{n}$ & ${+}\,\term{n}$ & ${+}\,\term{n}$ & ${+}\,{+}$ & ${+}\,($ & ${+}\,)$ & ${+}\,\term{w}$ & ${+}\,\term{n}\,{+}$ & ${+}\,)\,{+}$ \\
    $q_($   & $(\,\term{n}$ & $(\,\term{n}$ & $(\,\term{n}$ & $(\,{+}$ & $(\,($ & $(\,)$ & $(\,\term{w}$ & $(\,\term{n}\,{+}$ & $(\,)\,{+}$ \\
    \bottomrule
  \end{tabular}
  \caption{Token-spanner table for the running example of
  Figure~\ref{fig:greatgramma-example}: each cell lists the realizable sequence
  heads exposed by one token from one lexer state, for the full vocabulary.  The
  bold cell is one example of a stack-dependent head: a head containing
  \texttt{)} needs a matching \texttt{(} to pop.}
  \label{fig:token-spanner}
\end{figure}

\subsection{The Inverse Token-Spanner Table}

Everything so far is phrased in terms of terminal heads, but the checker must
ultimately mask \emph{tokens}.  The inverse token-spanner table bridges that gap by
reversing the relation above.  Given a realizable sequence head
\(T\mathbin{+\!\!+}[t]\) and the current lexer state \(q_a\), it returns the tokens
whose immediate output from \(q_a\) after reading the token is \(T\), and that
leave the final terminal \(t\) single-producible from the new lexer state.  In other words, it lists the tokens that begin a
parser-relevant terminal sequence from this lexer state, so once the checker knows
which heads the parser still accepts, the table turns them back into the tokens to
allow.

Inverting the \(q_N\) row of Figure~\ref{fig:token-spanner}, the head
\(\term{n}\,)\) is exposed from \(q_N\) only by \texttt{)}, so \((\term{n}\,),\ q_N)\) returns
\(\{\text{\texttt{)}}\}\).  The head \(\term{n}\) is exposed by \texttt{1}, \texttt{2},
and \texttt{12} (each only extends the number), so \((\term{n},\ q_N)\) returns
\(\{\text{\texttt{1}},\text{\texttt{2}},\text{\texttt{12}}\}\).

The construction \lean{BuildInverseTokenSpannerTable} computes both halves of the
table, the list of all realizable sequence heads and the inverse map.  The full
definition is in Appendix~\ref{app:lean-algorithm}; at a high level it walks every
lexer state and token, finds the single-producible terminals there with
\lean{computeSingleProducible} (Section~\ref{sec:correctness}), and records each
resulting head together with the tokens that expose it.

\subsection{Parser Preprocessing}

Online masking runs at every decoding step, so the goal is to precompute as much as
possible and leave as little online work as possible.  Parser preprocessing is the
key optimization on the parser side.  For each parser control state \(q_p\) and lexer
state \(q_a\), \lean{PreprocessParser} sorts the realizable sequence heads into three
classes.  The split rests on two facts about the parser, stack invariance and
over-approximation, both proved in Section~\ref{sec:correctness}.

\begin{enumerate}
  \item \textbf{Always allowed.}  The parser reads the head from the
  \emph{empty} stack and lands in a nonempty configuration set.  By stack
  invariance, readability is preserved under stack extension, so such a head is safe
  from any runtime stack, and its tokens are admitted with no online work.

  \item \textbf{Always rejected.}  The stack-forgetting NFA already rejects the
  head.  By over-approximation every parser run projects to an NFA run, so no
  stack can rescue it, and the head is dropped permanently.

  \item \textbf{Stack-dependent.}  Everything else.  Whether such a head is readable
  depends on the current stack and must be checked online.
\end{enumerate}

\subsection{Online Masking and the Assembled Checker}

At decoding time \lean{ComputeValidTokenMask} combines the preprocessed table with
the live parser stack.  It seeds the mask with the tokens of the always-allowed
heads, then tests each \emph{dependent} head against the live parser configuration
\((q_p, st)\), adding that head's tokens, recovered through the inverse
token-spanner table, when the head survives.  Always-rejected heads are never
touched as they are excluded from the dependent list in the offline stage.  In the listing, \lean{qa} is the current lexer state, \lean{qp} the parser
control state, and \lean{st} the parser stack.

\begin{lstlisting}
def ComputeValidTokenMask (P : PDA γ π σp)
    (itst : List γ → σa → List (Ch β)) (table : PPTable (Ch β) σp σa γ)
    (qa : σa) (qp : σp) (st : List π) : List (Ch β) :=
  -- tokens of the always-allowed heads
  let allowed   := (table qp qa).fst
  -- heads whose fate depends on stack st
  let dependent := (table qp qa).2.1
  (dependent.foldl
    (fun mask d =>
      -- keep iff d is readable from st
      if P.evalFrom {(qp, st)} d ≠ ∅
      then mask ++ itst d qa
      else mask)
      allowed).dedup
\end{lstlisting}

\lean{GCDChecker} evaluates the supplied prefix to obtain the lexer state and
reachable parser configurations, then unions their masks.  The extended alphabets
account for end-of-stream.  All correctness statements in
Section~\ref{sec:correctness} are about \lean{GCDChecker}.

All three classes of heads appear in the following example.  From the prefix
\(w=[\text{\texttt{1}}]\) (lexer in \(q_N\), parser in \(q_v\)), the candidate
\texttt{+} exposes the head \(\term{n}\,{+}\), which the parser reads from any stack without
failing, so it is always-allowed and admitted.  A token exposing \(\term{n}\,\term{n}\) is
always-rejected, since after a value the parser expects an operator, not another
value (the stack-forgetting NFA has no second-value edge from \(q_o\)).  The
candidate \texttt{)} exposes \(\term{n}\,)\), which is dependent, admissible only when
the stack holds a matching \texttt{(}, as after
\([\text{\texttt{(}},\text{\texttt{1}}]\) but not after \([\text{\texttt{1}}]\).

%% file: sections/correctness.tex
\section{Correctness}
\label{sec:correctness}

We now assemble the end-to-end theorem: we prove that the checker is \emph{sound} (every token
it allows from a viable prefix keeps the prefix viable) and \emph{complete} (every
viable token is allowed); together these say the prefixes the checker allows are
exactly the viable ones.  The proof has
three parts.  First, the precomputed tables compute what
Section~\ref{sec:algorithm} claimed.  Second, the parser preprocessing is
justified by two facts about pushdown acceptance.  Third, these combine into a
correct per-step decision that lifts to the language by induction.  Exact Lean
statements are in Appendix~\ref{app:lean}.

\subsection{The Precomputed Tables Are Correct}

Recall that the realizable sequence heads, and hence the whole token-spanner table
of Section~\ref{sec:algorithm}, are built from the single-producible terminals at
each lexer state.  Computing those, \lean{computeSingleProducible}, needs more than a structural induction, and it is the most technically
involved part of the table layer.  Finding the terminals the lexer can emit next
means chasing transitions that emit nothing until one emits a terminal, requiring a graph
search.  The search is the depth-first
traversal \lean{dfs} below, which follows empty-output transitions, records the
terminal on the first output-producing step of each path, and tracks a visited set
to avoid looping.

\begin{lstlisting}
def dfs (curr : σ) (vis : Finset σ)
    (ret : List γ) : List γ :=
  if curr ∈ vis then ret else
  let base := vis $\cup$ {curr}
  alphabet.foldl (init := ret)
    fun r c =>
    match M.step curr c with
    -- empty output: recurse
    | some (q', []) => dfs q' base r
    -- first terminal on this path
    | some (_, [t])  => t :: r
    -- no step, or multi-terminal output
    | _              => r
\end{lstlisting}

The search terminates because every recursive call adds a state to the visited set,
so the unvisited set strictly shrinks.  Correctness is two-sided.  \emph{Soundness}, that every recorded terminal is genuinely producible, is a
strong induction on the number of unvisited states.  An empty-output transition
appeals to the induction hypothesis at the next state, a single-terminal step is an
immediate witness, and multi-terminal steps are ignored because \lean{singleProducible}
records only single-terminal steps (multi-terminal output is captured separately by
the immediate-output component of a head).  \emph{Completeness}, that every producible
terminal is found, is the subtler converse.  The challenge is that a witness path
producing a single terminal may revisit a state, whereas \lean{dfs} visits each
state at most once.  To close this gap, we first decompose a run producing the
singleton output into an empty-output prefix, one transition producing the
terminal, and a suffix.  An empty-output cycle-removal lemma rewrites the prefix into a path with
no repeated states.  Such a path never trips the visited-set guard in \lean{dfs}, so it is
realized by \lean{DfsEpsReach}, a relation that records exactly the paths the
search follows.  An induction along that witness shows the search reaches the
state just before the terminal-producing transition.  The alphabet fold then
processes that transition, so the terminal appears in the result.

\begin{informaltheorem}[Tables match their specifications]
\label{thm:tables-exact}
The search \texttt{compute\allowbreak{}Single\allowbreak{}Producible} enumerates exactly the single-producible
terminals (\texttt{compute\allowbreak{}Single\allowbreak{}Producible\allowbreak{}\_correct}).  Consequently
\texttt{Build\allowbreak{}Inverse\allowbreak{}Token\allowbreak{}Spanner\allowbreak{}Table} equals its mathematical relation: its first
component lists exactly the realizable sequence heads (\lean{mem_re_iff}), and its
second returns, for each head and lexer state, exactly the tokens that begin
producing that head (\lean{mem_itst_iff}).
\end{informaltheorem}

Once the search is verified, the two table equalities follow quickly.

\subsection{The Parser Preprocessing Is Justified}

The three-way split of Section~\ref{sec:algorithm} rests on two operational facts
about the pushdown parser: stack invariance and over-approximation
(Appendix~\ref{app:lean-parser}).

\begin{informaltheorem}[Parser facts]
\label{thm:parser-facts}
\emph{Stack invariance}: if the parser can read a head from the empty stack, it can read
that head from every stack.  \emph{Over-approximation}: if the stack-forgetting NFA
has no run on a head, then the parser cannot read it from any stack.
\end{informaltheorem}

Both are inductions over the parser run, each resting on a single-step lemma.
\lean{stackInvariance} follows because a PDA move matches and rewrites only a
\emph{prefix} of the stack, so any deeper suffix is carried along untouched (the Lean
\lean{fullStep_stackInvariance}).  \lean{overApproximation} projects each PDA step
onto the stack-forgetting NFA, where dropping the stack only adds moves, so every
control state the PDA reaches is reachable in the NFA, and contrapositively an NFA
rejection forces a PDA rejection.  Together with Theorem~\ref{thm:tables-exact}, these
facts show the offline split of heads is sound: \lean{ComputeValidTokenMask} admits exactly the
tokens an unoptimized checker would, one with limited preprocessing that tests every head
against the live parser online.

\subsection{From the Decision to the Language}

It remains to connect the per-step mask to the language-level guarantee.  We prove
two statements about token prefixes.  Completeness says every viable prefix is
allowed, so every grammatical word is accepted.  Soundness says every prefix the
checker allows is still viable.

\paragraph{Completeness.}
At the language level, completeness says every word in the target text-lex-parse
language is accepted (\lean{TargetLanguage_imp_checkerAccepts}).  We prove it by
induction on the token prefix, showing at each step that the candidate is not masked.
Fix such a word and a prefix of it, and let \(P\) be the terminals emitted by the
prefix, \(S\) the terminals the next candidate token emits, and \(t\) the first
terminal of the remaining suffix, so the accepted terminal sequence begins
\(P \mathbin{+\!\!+} S \mathbin{+\!\!+} [t]\).  The head \(d := S \mathbin{+\!\!+} [t]\)
is a realizable sequence head whose inverse-table entry contains the candidate, and
\(d\) is a prefix of the accepted parse, so the mask keeps the candidate.
End-of-stream is handled separately as a special case.

\paragraph{Soundness.}
Soundness is the harder direction, and at the language level it takes the form of
\emph{productivity} (\lean{GCDChecker_productive}): every prefix the checker allows can
be extended to a complete accepted word.  Unwinding
how \lean{ComputeValidTokenMask} allows a token, via table correctness
(Theorem~\ref{thm:tables-exact}) and the parser facts (Theorem~\ref{thm:parser-facts}),
gives a realizable head the parser allows, namely
\(T\mathbin{+\!\!+}[t]\), with \(T\) produced by the token and \(t\)
single-producible afterwards.  Turning this into
a full completion is where the two pruning assumptions combine.  Parser prunedness gives
a \emph{terminal future}: from the parser configuration reached after this head, some
sequence of terminals leads to an accepting state.  Realizable tails
(Theorem~\ref{thm:realizable-tails}) handle the lexer side: a head ending in a
single-producible terminal can be continued by the lexer into any further terminal
sequence, up to whitespace that the parser ignores.  So the lexer can realize the
parser's terminal future, giving a complete accepted word that extends the prefix.

The two directions combine into the end-to-end theorem
\leanid{GCDChecker_intermediateLanguage_eq_TargetLanguage_prefixes},
which states that under the assumptions the prefixes the checker allows are exactly
the prefix closure of the target language.  We note that full correctness also requires
that \(\eos\) is masked correctly, which is comparatively simple and handled by
treating \(\eos\) as a special case.  The packaged theorem
\lean{GCDChecker_correct} includes a proof of correctness for handling \(\eos\) alongside the two directions above.

%% file: sections/discovered_issues.tex
\section{Formalization Findings}
\label{sec:findings}

Beyond confirming the algorithm, the formalization surfaced (1) a concrete
bug in the published lexing construction, and (2) a precise account of the
assumptions the algorithm relies on.  We describe both, then the artifact that
instantiates the whole pipeline.

\subsection{The Proof Uncovered a Lexing Bug}

While proving the equivalence between the executable lexing transducer
\lean{BuildLexingFST} and the relational lexer specification \lean{PartialLexRel}
(Theorem~\ref{thm:lexing-equiv}), we found that the published construction does
not always lex correctly.  In the \greatgramma paper the lexing transducer reuses
the state space of the lexer automaton, and, so that end-of-stream (\(\eos\)) can
be accepted as the very first symbol, it adds an \(\eos\) self-loop at the
automaton's start state \(q_0\).  This is safe when the automaton never returns to
\(q_0\).  When the automaton does return, the self-loop accepts \(\eos\) in the
middle of an unfinished terminal.

\begin{figure}[t]
  \centering
  \begin{minipage}{0.46\linewidth}\centering
  \begin{tikzpicture}[>=stealth,shorten >=1pt,node distance=2.0cm,on grid,auto,every node/.style={font=\small},every state/.style={minimum size=15pt,inner sep=1pt}]
    \node[state,initial below] (q0) {$q_0$};
    \node[state,accepting,right=of q0] (q1) {$q_1$};
    \path[->]
      (q0) edge[bend left] node {a} (q1)
      (q1) edge[bend left] node[below] {b} (q0);
  \end{tikzpicture}\\[2pt]
  {\footnotesize (a) automaton for $(ab)^*a$}
  \end{minipage}\hfill
  \begin{minipage}{0.50\linewidth}\centering
  \begin{tikzpicture}[>=stealth,shorten >=1pt,node distance=2.2cm,on grid,auto,every node/.style={font=\small},every state/.style={minimum size=15pt,inner sep=1pt}]
    \node[state,initial below] (q0) {$q_0$};
    \node[state,right=of q0] (q1) {$q_1$};
    \path[->]
      (q0) edge[bend left] node {a : $\epsilon$} (q1)
      (q1) edge[bend left] node[below] {b : $\epsilon$} (q0)
      (q0) edge[loop above] node {$\eos$ : $\eos$} ()
      (q1) edge[loop above] node {a : T} ();
  \end{tikzpicture}\\[2pt]
  {\footnotesize (b) the buggy FST construction}
  \end{minipage}
  \caption{The lexing bug.  Reading \texttt{b} from the accepting state \(q_1\) of
  the automaton (a) returns to \(q_0\).  The published FST (b) reuses these states
  and adds an \(\eos\) self-loop at \(q_0\) (other emitting transitions omitted), so
  \texttt{a}\,\texttt{b}\,\(\eos\) is accepted even though \texttt{ab} is not a
  complete terminal.}
  \label{fig:abstar-a}
\end{figure}

The regular expression \((ab)^*a\) triggers this case
(Figure~\ref{fig:abstar-a}).  The fix gives the transducer a distinguished start
state instead of reusing \(q_0\), so accepting \(\eos\) at the very beginning no
longer leaks into mid-terminal positions.  This start state is what
\lean{BuildLexingFST} uses throughout (Section~\ref{sec:lexing}), and it is what
makes Theorem~\ref{thm:lexing-equiv} hold.

This is representative of the kind of boundary mistakes the GCD literature keeps hitting.  In
their evaluation, \citet{park2025flexible} found that decoders from prior published
work mask tokens that should not be masked and, on some simple grammars, fail to
terminate during preprocessing. For instance, XGrammar~\cite{dong2024xgrammar} could not be run at
all on their benchmarks, failing to compile one grammar and looping or segfaulting
during decoding on others.  These are carefully engineered systems,
yet the token--terminal boundary still trips them up.  Our bug is in the same
family and survived a principled construction. A machine-checked
proof, as opposed to human review and testing, is what surfaced it.

\subsection{The Assumptions}
\label{subsec:findings-separator}

Stating the theorem in Lean made the algorithm's assumptions concrete where the
paper had left them implicit.  The end-to-end result holds under a small bundle
(Section~\ref{subsec:assumptions}): the lexer recognizes no empty terminal, every
character is itself a token, and the lexer and parser have no dead states.  The one
that does real work is the \emph{universal separator}, a distinguished whitespace
terminal that ends any lexeme and is a no-op for the parser.  Pinning it down
precisely is what lets the realizable-tail characterization collapse the otherwise
unbounded set of lexer continuations to a finite check
(Theorem~\ref{thm:realizable-tails}).  Our JSON-like instance satisfies this
assumption using newline as the separator, although other grammars require their own proofs.

\subsection{Artifact and a Worked Example}

The artifact is a Lean~4 development (see the introduction for the repository).
Appendix~\ref{app:theorem-map} records the module map, the theorem dependencies, and
build notes.

We emphasize that our correctness theorem is generic.  It holds for every
vocabulary, lexer, and parser that satisfy the stated assumptions.  To demonstrate
that the assumptions can be proved in practice, we carry out the following
experiment.  We instantiate the whole pipeline on a concrete grammar and prove the
assumptions for it.  \lean{GCDTest.lean} defines a JSON character
and terminal alphabet, a lexer for braces, brackets, strings, numbers, booleans,
null, punctuation, and newline whitespace, a pushdown parser for a shallow JSON
language with primitive values and arrays and objects, and a vocabulary mixing
singleton tokens with composite ones such as \texttt{true}, \texttt{false},
\texttt{null}, string literals, digits, and blank lines.  It establishes all of the
\lean{GCDAssumptions}, including the universal separator and both prunedness
conditions, and derives the correctness theorems, yielding
\leanid{jsonChecker_correct}, the final theorem specialized to this grammar.

%% file: sections/related_work.tex
\section{Related Work}
\label{sec:related-work}

We place this work in four areas: constrained decoding, verified parsing,
verified lexing, and Lean's mechanized mathematics infrastructure.  The first
three provide the algorithmic context and reusable proof ideas; the last helps
explain which parts of the development rely on general-purpose libraries.

\subsection{Constrained Decoding for Language Models}

Grammar-constrained and guided decoding methods restrict next-token choices to
make model outputs structurally valid, spanning structured NLP~\cite{geng2023gcd},
semantic decoding for code~\cite{poesia2022synchromesh}, regular-expression and CFG
constraints~\cite{willard2023efficient}, tokenizer/constraint
alignment~\cite{beurerkellner2024domino}, and efficient CFG-oriented
systems~\cite{ugare2024syncode,dong2024xgrammar}.  They differ in expressiveness,
preprocessing, tokenizer integration, and latency, and those choices are where
soundness and completeness bugs hide (Section~\ref{sec:findings}); such bugs in
carefully engineered, peer-reviewed systems motivate a machine-checked proof.

The algorithm formalized here is \greatgramma by Park et
al.~\cite{park2025flexible}.  \greatgramma introduces the inverse token-spanner
and parser-preprocessing architecture mechanized in this paper, and subsequent work
has already reused its lexical preprocessing~\cite{li2026psc}.  We provide a
complementary contribution: a machine-checked correctness theorem for the core
automata construction, together with an explicit account of the assumptions under
which it is exact.

\subsection{Verified Parsing}

Verified parsing is a natural comparison point.  Prior work verifies packrat
parsers for PEGs~\cite{blaudeau2020packrat}, validates LR(1) parsers in
Coq~\cite{jourdan2012lr1}, and verifies an LL(1) parser generator~\cite{lasser2019ll1},
an ALL(*) parser~\cite{lasser2021costar},
linear-time LL(1) derivative parsing~\cite{edelmann2020zippy}, and an Earley parser
in Isabelle/HOL~\cite{rau2024earley}.  These recognize or produce parse results for
strings, and suggest the pattern we follow: define a semantic reference, define the
executable object, and prove them equal.  We instead assume a PDA-style parser and
verify a next-token mask around it, where the central challenge is preserving viable
parser continuations across detokenization, lexing, token boundaries, and stacks.

\subsection{Verified Lexing and Regular-Expression Matching}

Verified lexing and regular-expression matching provide further precedent.
Verbatim++ verifies optimized derivative-based lexing with semantic
actions~\cite{egolf2022verbatimpp}.  Chattopadhyay, Li, and Mamouras verify an
efficient matcher for regular expressions with
lookaround~\cite{chattopadhyay2025lookaround}.  Like our development, their 
formalization relates an algorithm to standard matching semantics.  Coquand and
Siles formalize regular-expression equivalence in type theory~\cite{coquand2011regexp}.
Our lexer proof is driven by constrained decoding: the lexer is an FST composed with
detokenization and queried at token boundaries, and the key theorem equates partial
lexing with the FST states and outputs exposed to the checker.

\subsection{Lean and Mechanized Mathematics Infrastructure}

The development is written in Lean~4~\cite{demoura2021lean4} and uses mathlib
infrastructure for finite sets, lists, automata-related constructions, and proof
automation~\cite{mathlib2020}.  Partial FSAs
and FSTs with executable composition (\lean{Automata.lean}), pushdown automata with
their stack lemmas (\lean{PDA.lean}), and the verified graph search
(\lean{Producible.lean}) are artifact modules with no GCD-specific dependencies and
can be imported by other Lean developments as they stand.

%% file: sections/limitations_conclusion.tex
\section{Discussion}
\label{sec:limitations}

\subsection{Formalization Lessons}
\label{subsec:design-lessons}

The development suggests a reusable approach: give each interface, from tokens
through characters and terminals to parser configurations, an executable
implementation and a relational or language-level specification, then prove the
two equivalent.  The end-to-end proof can use whichever representation suits each
argument.  For example, the lexer runs as a transducer, while its relational
specification states maximal-munch behavior directly.  This separation helped
isolate the end-of-stream defect.

The bridge proofs add work, and component changes may require revisiting them. We
chose this structure because the optimized checker is efficient to run, while its
semantic specification is easier to reason about compositionally. The theorem
states the assumptions at component boundaries; it verifies their composition,
but does not derive a lexer and parser from a grammar description.

Computing the single-producible terminals at a lexer state is technically demanding
to verify.  Showing that a depth-first search with a visited set enumerates them
exactly takes more than a structural induction (Section~\ref{sec:correctness}).
The techniques it uses, including an explicit reachability specification, a
soundness induction on the unvisited set, and a cycle-removal lemma showing that
visited-set pruning loses no witness, carry over to other nontrivial graph
algorithms.

AI assistance helped with connective proof work, but less with the central
conceptual arguments. We identified the universal-separator assumption and
developed and proved the exchange arguments at the heart of the paper in
hand-written Lean. After establishing the local lemmas and interfaces, we used an
AI assistant to help assemble them into the full soundness and completeness proof
and suggest intermediate statements. We checked each suggestion in Lean.

\subsection{Scope and Limitations}

The theorems are parametric: each grammar supplies a lexer, parser, vocabulary,
and proofs of \lean{GCDAssumptions}; the generic correctness results need not be
reproved. The universal-separator condition limits applicability, especially for
whitespace-sensitive languages (Section~\ref{subsec:assumptions}). Our detokenizer
uses a different construction from \greatgramma's, but an equivalence theorem
shows that both produce the same language (Section~\ref{sec:lexing}). Our
single-producible search is exponential and prioritizes proof clarity over runtime.

\section{Conclusion}
\label{sec:conclusion}

Our Lean formalization proves the checker sound and complete under explicit
assumptions, and exposes a lexer bug and a previously implicit separator
condition. It also illustrates a reusable method: specify interfaces
semantically, verify executable summaries against those specifications, and state
the assumptions in the end-to-end theorem.

%% file: sections/data_availability.tex
\section*{Data-Availability Statement}

The Lean~4 development that accompanies this paper is archived on Zenodo~\cite{bhat2026artifact} under DOI 10.5281/zenodo.24447388. It is also available at \url{https://github.com/large-loris-models/great-gramma-formalization}, and the version evaluated in this paper is commit 989fd77dc39b7d2e97bee8719a797efa3afac9fc.

%% file: sections/appendix_lean.tex
\section{Lean Definitions and Theorems}
\label{app:lean}

\lstset{basicstyle=\footnotesize\ttfamily,breakatwhitespace=false,
  escapeinside={(*@}{@*)}}

The body of the paper states most definitions and theorems in prose, keeping
only the Lean that clarifies how an object is modeled.  This appendix collects
the executable definitions and exact Lean statements behind those prose theorems,
so that a reader can match each claim to its mechanized counterpart.  The names
are Lean-facing and can be searched directly in the artifact.

\subsection{Verified Lexing}
\label{app:lean-lexing}

The one-lookahead lexer is specified as the inductive relation
\lean{PartialLexRel}, whose constructors are the cases listed in
Section~\ref{sec:lexing}.

\begin{lstlisting}
-- Lex(characters) = (terms, unlexed)
-- PartialLexRel characters terms unlexed
inductive PartialLexRel (spec : LexerSpec α γ σ)
  : List (Ch α) -> List (Ch γ) -> List α -> Prop
| nil : PartialLexRel spec [] [] []
| step_nil_eos {w wn terms} :
    PartialLexRel spec w terms [] ->
    wn = w ++ [ExtChar.eos] ->
    [] ∉ spec.automaton.accepts ->
    PartialLexRel spec wn (terms ++ [.eos]) []
| step_eos {w wn terms unlexed} :
    PartialLexRel spec w terms unlexed ->
    wn = w ++ [ExtChar.eos] ->
    (h : unlexed ∈ spec.automaton.accepts) ->
    PartialLexRel spec wn
      (terms ++ [.char (spec.accept_seq_term unlexed h), .eos]) []
| step_char_continue {w wn terms unlexed ch} :
    PartialLexRel spec w terms unlexed ->
    wn = w ++ [ExtChar.char ch] ->
    unlexed ++ [ch] ∈ spec.automaton.prefixLanguage ->
    PartialLexRel spec wn terms (unlexed ++ [ch])
| step_char_commit {w wn terms unlexed ch} :
    PartialLexRel spec w terms unlexed ->
    wn = w ++ [ExtChar.char ch] ->
    unlexed ++ [ch] ∉ spec.automaton.prefixLanguage ->
    [ch] ∈ spec.automaton.prefixLanguage ->
    (h : unlexed ∈ spec.automaton.accepts) ->
    PartialLexRel spec wn
      (terms ++ [ExtChar.char (spec.accept_seq_term unlexed h)]) [ch]
\end{lstlisting}

The executable lexing transducer \lean{BuildLexingFST} implements
\greatgramma's construction (Algorithm 2 of Park et al.~\cite{park2025flexible}).

\begin{lstlisting}
def BuildLexingFST (spec : LexerSpec α Γ σ)
  : FST (Ch α) (Ch Γ) (LexingState σ) := Id.run do
  let (*@$\langle$@*)A, term, hterm, _, _(*@$\rangle$@*) := spec
  let new_q0 := LexingState.start
  let q0 := A.start
  let F := A.accept
  let step := fun (q : LexingState σ) (c : Ch α) =>
    let qorg := LexingState.src spec q
    match c with
    | ExtChar.char c =>
      if h : (A.step qorg c).isSome then
        some (LexingState.id ((A.step qorg c).get h), [])
      else if h : qorg ∈ F then
        let T := (term qorg).get <| ((hterm qorg).mp h)
        A.step q0 c |>.map fun q' => (LexingState.id q', [ExtChar.char T])
      else
        none
    | ExtChar.eos =>
      if h : qorg ∈ F then
        let T := (term qorg).get <| (hterm qorg).mp h
        some (new_q0, [T, .eos])
      else if q = new_q0 then
        some (new_q0, [.eos])
      else
        none
  (*@$\langle$@*)new_q0, step, {new_q0}(*@$\rangle$@*)
\end{lstlisting}

The two transducers agree (Section~\ref{sec:lexing}):

\begin{lstlisting}
theorem PartialLexRel_to_LexingFST (spec: LexerSpec α Γ σ) (he: [] ∉ spec.automaton.accepts) (hpruned: spec.automaton.pruned) :
  ∀ w terms unlexed,
    PartialLexRel spec w terms unlexed →
    ∃ q', (BuildLexingFST spec).eval w = some (q', terms) ∧
      (BuildLexingFST spec).eval unlexed = some (q', [])

theorem LexingFST_to_PartialLexRel (spec: LexerSpec α Γ σ) (he: [] ∉ spec.automaton.accepts) (hpruned: spec.automaton.pruned) :
  ∀ w q' terms,
    (BuildLexingFST spec).eval w = some (q', terms) →
    ∃ (unlexed : List α),
      (BuildLexingFST spec).eval unlexed = some (q', []) ∧ PartialLexRel spec w terms unlexed
\end{lstlisting}

The realizable-sequence characterization (Section~\ref{subsec:realizable-sequences})
says that, modulo whitespace, the realizable tails from a lexer state are exactly
those beginning with a \leanid{singleProducible} terminal.

\begin{lstlisting}[basicstyle=\scriptsize\ttfamily]
lemma tailModdedRealizableSequences_eq_singleProducibleHead
  (spec: LexerSpec α Γ σ)
  (hempty : [] ∉ spec.automaton.accepts)
  (hpruned: spec.automaton.pruned)
  (hwa: WhitespaceAssumption spec tnonwhite twhite qnonwhite qwhite) (q: LexingState σ) :
  let white_term := (whitespaceTerminal spec tnonwhite twhite qnonwhite qwhite hwa)
  let lexer := BuildDetokLexer (v := vocab) spec
  lexer.tailModdedRealizableSequences (Unit.unit, q) white_term =
    { Ts | Ts = [] ∨ (∃ t tsfx,
          ¬Ts.contains (ExtChar.char white_term) ∧
          Ts = t :: tsfx ∧ t ∈ lexer.singleProducible (Unit.unit, q)) }
\end{lstlisting}

The converse of the realizable-tail characterization is assembled from four
exchange lemmas (Section~\ref{subsec:realizable-sequences}), all
\lean{private} in \lean{Lexing/Whitespace.lean}.  Each starts from a state where
some terminal is single-producible and re-derives it by a run that ends in the
clean post-separator state \lean{qwhite}, so that the runs for successive target
terminals can be concatenated.  In the statements below, \lean{lexer} abbreviates
\lean{BuildDetokLexer spec} and \lean{white_term} the whitespace terminal
\lean{whitespaceTerminal spec tnonwhite twhite qnonwhite qwhite hwa}.

\begin{lstlisting}[basicstyle=\scriptsize\ttfamily]
-- a single-producible non-whitespace terminal, re-produced ending at qwhite
private lemma exchangeNonWhitespace (spec : LexerSpec α Γ σ)
  (hempty : [] ∉ spec.automaton.accepts)
  (hwa : WhitespaceAssumption spec tnonwhite twhite qnonwhite qwhite)
  (q : LexingState σ) (term : Γ)
  (hterm : term ≠ white_term ∧
           ExtChar.char term ∈ lexer.singleProducible (Unit.unit, q)) :
  ∃ w, lexer.evalFrom (Unit.unit, q) w
         = some ((Unit.unit, LexingState.id qwhite), [ExtChar.char term])

-- likewise for eos
private lemma exchangeEOS (spec : LexerSpec α Γ σ)
  (hwa : WhitespaceAssumption spec tnonwhite twhite qnonwhite qwhite)
  (q : LexingState σ)
  (hterm : ExtChar.eos ∈ lexer.singleProducible (Unit.unit, q)) :
  ∃ w, lexer.evalFrom (Unit.unit, q) w
         = some ((Unit.unit, LexingState.id qwhite), [ExtChar.eos])

-- from a state where white_term is single-producible, emit it while landing in
-- any state other than the start and whitespace states
private lemma exchangeWhitespace (spec : LexerSpec α Γ σ)
  (hempty : [] ∉ spec.automaton.accepts) (hpruned : spec.automaton.pruned)
  (hwa : WhitespaceAssumption spec tnonwhite twhite qnonwhite qwhite)
  (q : LexingState σ)
  (hwsa : ExtChar.char white_term ∈ lexer.singleProducible (Unit.unit, q)) :
  ∀ qtarget, qtarget ≠ qwhite → qtarget ≠ spec.automaton.start →
    ∃ w, lexer.evalFrom (Unit.unit, q) w
           = some ((Unit.unit, LexingState.id qtarget),
                   [ExtChar.char white_term])

-- the separator followed by eos, ending at qwhite
private lemma exchangeWhitespaceEOS (spec : LexerSpec α Γ σ)
  (hempty : [] ∉ spec.automaton.accepts)
  (hwa : WhitespaceAssumption spec tnonwhite twhite qnonwhite qwhite)
  (q : LexingState σ)
  (hwsa : ExtChar.char white_term ∈ lexer.singleProducible (Unit.unit, q)) :
  ∃ w, lexer.evalFrom (Unit.unit, q) w
         = some ((Unit.unit, LexingState.id qwhite),
                 [ExtChar.char white_term, ExtChar.eos])
\end{lstlisting}

\subsection{Inverse Token-Spanner Table, Preprocessing, and Checker}
\label{app:lean-algorithm}

The inverse token-spanner table is computed by
\leanid{BuildInverseTokenSpannerTable} (Section~\ref{sec:algorithm}).

\begin{lstlisting}
variable [q : FinEnum σf] [a : FinEnum α] [t : FinEnum Γ]

def BuildInverseTokenSpannerTable
  (fst_comp : FST α Γ σf) : Re Γ × (List Γ → σf → (List α)) := Id.run do
  let states := q.toList
  let A := a.toList

  let re :=
    states.flatMap (fun q =>
      A.flatMap (fun c =>
        match fst_comp.step q c with
        | none => []
        | some (q', Ts) =>
          (fst_comp.computeSingleProducible q')
          |>.map (fun t => Ts ++ [t])
      )
    )
    |>.eraseDups

  let tinv := fun rs s =>
    if h : rs ≠ [] then
      let Ts := rs.dropLast
      let T := rs.getLast h
      A.filter (fun c =>
        match fst_comp.step s c with
        | none => false
        | some (q', Ts') =>
            (fst_comp.computeSingleProducible q').contains T && Ts' = Ts)
    else []

  (re, tinv)
\end{lstlisting}

The online mask and the assembled checker are
\leanid{ComputeValidTokenMask}, \lean{MaskChecker}, and the \greatgramma
specialization \lean{GCDChecker}.

\begin{lstlisting}
def ComputeValidTokenMask (P : PDA γ π σ_p)
    (itst : List γ → σ_a → List β)
    (table : PPTable β σ_p σ_a γ)
    (qa : σ_a) (qp : σ_p) (st : List π) : List β :=
  let allowed := (table qp qa).fst
  let dependent := (table qp qa).2.1
  let allowed :=
    dependent.foldl
      (fun acc d => if P.evalFrom {(qp, st)} d ≠ ∅
                    then acc ++ itst d qa else acc)
      allowed
  allowed.dedup

def MaskChecker
  (comb : FST (Ch β) (Ch Γ) σa) (parser : PDA (Ch Γ) π σp)
  (pp_table : PPTable (Ch β) σp σa (Ch Γ))
  (itst : List (Ch Γ) → σa → List (Ch β))
  : List β → Ch β → Bool :=
  fun curr cand =>
    match comb.eval (curr.map ExtChar.char) with
    | none => false
    | some (q_fst, terms) =>
      let q_pda := parser.evalFrom {(parser.start, [])} terms
      let in_curr := q_pda.image
        (fun (q_parse, st) => (ComputeValidTokenMask parser itst pp_table q_fst q_parse st).contains cand)
      Finset.fold Bool.or false id in_curr

GCDChecker spec P =
  MaskChecker
    (Detokenizing.BuildDetokLexer (V := Ch β) spec)
    (ParserWithEOS P)
    (PreprocessParser (Detokenizing.BuildDetokLexer (V := Ch β) spec) (ParserWithEOS P))
    (BuildInverseTokenSpannerTable (Detokenizing.BuildDetokLexer (V := Ch β) spec)).snd
\end{lstlisting}

The computed table matches its mathematical relation
(Section~\ref{sec:correctness}):

\begin{lstlisting}
lemma mem_re_iff
  (fst_comp : FST α Γ σf) (d : List Γ) :
  d ∈ (BuildInverseTokenSpannerTable fst_comp).fst ↔
  d ∈ RealizableSequenceHeads fst_comp

lemma mem_itst_iff
  (fst_comp : FST α Γ σf) (d : List Γ) (qa : σf) (tok : α) :
  tok ∈ (BuildInverseTokenSpannerTable fst_comp).snd d qa ↔ tok ∈ InverseTokenSpannerTable fst_comp d qa
\end{lstlisting}

\subsection{Parser Facts}
\label{app:lean-parser}

The two parser lemmas justifying the offline partition
(Section~\ref{sec:correctness}):

\begin{lstlisting}
/-- Acceptance is monotone under extension of the initial stack. -/
theorem stackInvariance : ∀ w s st st',
  st <+: st' → w ∈ P.acceptsFrom s st → w ∈ P.acceptsFrom s st'

theorem overApproximation : ∀ w, w ∉ P.toNFA.accepts → w ∉ P.accepts
\end{lstlisting}

\subsection{Single-Output Search}
\label{app:lean-dfs}

The single-producible terminals are enumerated by the depth-first search
\lean{dfs} (Section~\ref{sec:correctness}).

\begin{lstlisting}
def dfs
  [ Fintype Γ ] [ Fintype σ ] [ a: FinEnum α ]
  [DecidableEq σ ] [DecidableEq α ] [DecidableEq Γ ]
  (curr : σ) (vis : Finset σ) (ret : List Γ) : List Γ :=
  let alph := a.toList
  if _h_vis : curr ∈ vis then ret
  else
    let base := vis (*@$\cup$@*) {curr}
    let stepAccum : List Γ → α → List Γ :=
      fun retacc next =>
        match M.step curr next with
        | some (nextState, []) => dfs nextState base retacc
        | some (_, [γ]) => γ :: retacc
        | _ => retacc
    alph.foldl stepAccum ret
  termination_by (vis(*@$^{c}$@*)).card
  decreasing_by
    exact compl_card_lt_of_insert _h_vis
\end{lstlisting}

\subsection{End-to-End Theorems}
\label{app:lean-e2e}

The completeness direction, the complete-language equality, and the
prefix-language equality that together express soundness and completeness of the
checker (Section~\ref{sec:correctness}):

\begin{lstlisting}[basicstyle=\scriptsize\ttfamily]
theorem TargetLanguage_imp_checkerAccepts
  (spec : LexerSpec α Γ σa) (P : PDA Γ π σp)
  {tnonwhite twhite : α} {qnonwhite qwhite : σa}
  (hassum : GCDAssumptions spec P tnonwhite twhite qnonwhite qwhite)
  (w : List β) (hw : w ∈ TargetLanguage spec P) :
  checkerAccepts (GCDChecker spec P) w = true

theorem GCDChecker_checkerLanguage_eq_TargetLanguage
  (spec : LexerSpec α Γ σa) (P : PDA Γ π σp)
  {tnonwhite twhite : α} {qnonwhite qwhite : σa}
  (hassum : GCDAssumptions spec P tnonwhite twhite qnonwhite qwhite) :
  checkerLanguage (β := β) (GCDChecker spec P) = TargetLanguage spec P

theorem GCDChecker_intermediateLanguage_eq_TargetLanguage_prefixes
  (spec : LexerSpec α Γ σa) (P : PDA Γ π σp)
  {tnonwhite twhite : α} {qnonwhite qwhite : σa}
  (hassum : GCDAssumptions spec P tnonwhite twhite qnonwhite qwhite) :
  checkerIntermediateLanguage (β := β) (GCDChecker spec P) =
    (TargetLanguage (β := β) spec P).prefixes
\end{lstlisting}

%% file: sections/appendix_theorem_map.tex
\section{Theorem Map}
\label{app:theorem-map}

This appendix records the main Lean definitions and theorems, with names that
can be searched directly in the artifact.  We first list the definitions that
anchor the development, then map the paper's algorithms and results to Lean
declarations, and finish with notes about the artifact.

\subsection{Definitions Worth Looking Up First}

The most important executable objects appear in a short chain:
\begin{lstlisting}
PartialLexRel spec w tokens unlexed
PartialLex spec w = some (tokens, unlexed)
BuildLexingFST spec : FST (Ch α) (Ch γ) (LexingState σ)
BuildDetokLexer spec =
  FST.compose BuildDetokenizingFST (BuildLexingFST spec)
\end{lstlisting}
These are not cosmetic variants of the same object.
\lean{PartialLexRel} is the readable semantic specification, \lean{PartialLex} is the direct executable partial lexer, and \lean{BuildLexingFST} is the finite transducer used inside the checker.  The correctness development proves that they agree extensionally, so later proofs can move to whichever presentation is most convenient.

The detokenization-to-preprocessing bridge also has a concrete Lean shape.  The set of realizable sequence heads is:
\begin{lstlisting}
def RealizableSequenceHeads (fst_comp : FST α γ σf) :
  Set (List γ) :=
  { Ts' | ∃ q0 t Ts q1 T,
      fst_comp.step q0 t = some (q1, Ts) ∧
      T ∈ fst_comp.singleProducible q1 ∧
      Ts' = Ts ++ [T] }
\end{lstlisting}
The executable object consumed by preprocessing is \leanid{BuildInverseTokenSpannerTable}, and the key equivalence theorems are \lean{mem_re_iff} and \lean{mem_itst_iff}.

\subsection{Detokenization and Prefix-Exposure Lemmas}

The proof-critical detokenization facts are easiest to read as a forward/backward chain:
\begingroup\emergencystretch=4em\relax
\begin{itemize}
  \item \lean{detokenizer_comp}: the executable detokenizing FST is equivalent to flattening first and then running the downstream FST;
  \item \leanid{BuildDetokLexer_singleProducible_of_evalFrom}: if the composed detokenizing lexer produces a nonempty output sequence from a state, then the head output symbol is single-producible from that state;
  \item \leanid{tailModdedRealizableSequences_eq_singleProducibleHead}: after quotienting away harmless separator insertion, realizable tails are characterized exactly by their single-producible head.
\end{itemize}
\endgroup
These are the lemmas that let the paper move from an arbitrary semantic continuation to the finite witness actually used by preprocessing.  In other words, this is the theorem family that turns ``there is a good continuation'' into ``the checker can see a finite one-step reason to allow this token.''

\begin{table*}[!t]
\caption{Main theorem dependencies.}
\label{tab:theorem-map}
\scriptsize
\renewcommand{\arraystretch}{0.95}
\begin{tabularx}{\textwidth}{@{}p{0.16\textwidth}p{0.44\textwidth}X@{}}
\toprule
Layer & Lean artifact & Informal role \\
\midrule
Generic checker & \lean{checkerCorrect} & language equality plus prefix equality \\
FSA semantics & \lean{FSA.pruned_prefixLanguage} & pruned FSA intermediate language equals prefix language \\
PDA semantics & \lean{PDA.stackInvariance} & acceptance monotone under stack extension \\
PDA semantics & \lean{PDA.overApproximation} & stack-forgetting NFA rejection implies PDA rejection \\
Lexer relation & \lean{PartialLex_pruned_eq_PartialLexRel} & executable partial lexer equals relational partial lexer \\
Lexer FST & \lean{PartialLex_to_LexingFST} & executable lexer implies FST run \\
Lexer FST & \lean{PartialLexRel_to_LexingFST} & relational lexer implies FST run \\
Lexer FST & \lean{LexingFST_to_PartialLexRel} & FST run implies relational lexer \\
Detokenizer & \lean{detokenizerFST_eq_detokenizer} & detokenizing FST equals list flattening \\
Detokenizer & \lean{detokenizer_comp} & composed FST equals detokenize-then-run \\
Detok bridge & \lean{BuildDetokLexer_singleProducible_of_evalFrom} & nonempty detok-lexer run exposes single-producible head \\
Detok bridge & \lean{tailModdedRealizableSequences_eq_singleProducibleHead} & separator-normalized tails equal single-producible-head characterization \\
Single outputs & \lean{computeSingleProducible_correct} & DFS enumerates exactly single-producible outputs \\
Realizable sequence heads & \lean{mem_re_iff} & executable head list equals the semantic set \\
Token spanner & \lean{mem_itst_iff} & executable inverse token-spanner table equals semantic relation \\
Mask & \lean{mem_ComputeValidTokenMask_preprocess_iff} & exact membership characterization for online mask \\
Step theorem & \lean{Soundness} & allowed token implies viable semantic step \\
Step theorem & \lean{Completeness} & viable semantic step implies allowed token \\
EOS theorem & \lean{EOSCompleteness} & viable EOS step implies EOS is allowed \\
Language theorem & \leanid{GCDChecker_checkerLanguage_eq_TargetLanguage} & accepted checker words equal semantic target language \\
Prefix theorem & \leanid{GCDChecker_intermediateLanguage_eq_TargetLanguage_prefixes} & allowed prefixes equal semantic prefixes \\
Soundness & \lean{GCDChecker_productive} & every allowed prefix has an accepted extension \\
Path independence & \lean{GCDChecker_pathIndependent} & checker depends only on flattened character content \\
Final theorem & \lean{GCDChecker_correct} & packaged checker correctness theorem \\
JSON instance & \lean{jsonChecker_correct} & final theorem instantiated for the shallow JSON example \\
\bottomrule
\end{tabularx}
\end{table*}

\subsection{Artifact Module Map}

Table~\ref{tab:artifact-map} lists the main modules and theorem families in the Lean development.

\begin{table*}[!t]
\caption{Lean module map.}
\label{tab:artifact-map}
\small
\renewcommand{\arraystretch}{0.95}
\setlength{\tabcolsep}{5pt}
\begin{tabularx}{\textwidth}{@{}p{0.24\textwidth}p{0.20\textwidth}X@{}}
\toprule
Module & Role & Representative definitions/theorems \\
\midrule
\lean{Char.lean} & EOS-extended symbols & \lean{ExtChar}, \lean{Ch} \\
\lean{Vocabulary.lean} & token-to-character interface & \lean{Vocabulary.flatten}, \lean{Vocabulary.embed} \\
\lean{Automata.lean} & deterministic FSAs/FSTs & \lean{FSA.pruned_prefixLanguage}, \lean{FST.compose_correct} \\
\lean{PDA.lean} & pushdown parser semantics & \lean{PDA.stackInvariance}, \lean{PDA.overApproximation} \\
\lean{Lexing/Base.lean} & lexer specification and FST construction & \lean{LexerSpec}, \lean{PartialLexRel}, \lean{BuildLexingFST} \\
\lean{Lexing/Correctness.lean} & lexer equivalence proofs & \lean{PartialLex_pruned_eq_PartialLexRel}, \lean{LexingFST_to_PartialLexRel} \\
\lean{Lexing/Detokenizing.lean} & detokenizer and FST composition & \lean{detokenizerFST_eq_detokenizer}, \lean{detokenizer_comp} \\
\lean{Lexing/Whitespace.lean} & whitespace exchange & \lean{BuildDetokLexer_singleProducible_of_evalFrom}, \lean{tailModdedRealizableSequences_eq_singleProducibleHead} \\
\lean{Producible.lean} & single-output DFS & \lean{computeSingleProducible_correct} \\
\lean{RealizableSequence.lean} & inverse token-spanner tables & \lean{RealizableSequenceHeads}, \lean{mem_re_iff}, \lean{mem_itst_iff} \\
\lean{ParserWithEOS.lean} & parser EOS adapter & \lean{ParserWithEOS} \\
\lean{GCDAssumptions.lean} & global assumptions & \lean{GCDAssumptions}, \lean{GCDWhitespaceAssumption} \\
\lean{GCDAlgorithm.lean} & executable preprocessing and checker & \lean{PreprocessParser}, \lean{ComputeValidTokenMask}, \lean{GCDChecker} \\
\lean{GCDStepProofs.lean} & step-level mask correctness & \lean{Soundness}, \lean{Completeness}, \lean{EOSCompleteness} \\
\lean{GCDCheckerLanguage.lean} & semantic language equality & \lean{TargetLanguage}, \leanid{GCDChecker_checkerLanguage_eq_TargetLanguage} \\
\lean{GCDProductivity.lean} & soundness and final theorem & \lean{GCDChecker_productive}, \lean{GCDChecker_correct} \\
\lean{GCDTest.lean} & finite JSON-like worked example & \lean{jsonChecker_correct}, \lean{jsonChecker_pathIndependent} \\
\bottomrule
\end{tabularx}
\end{table*}

\subsection{Artifact Notes}

The repository is organized by proof layer, from basic automata and vocabularies through lexing, preprocessing, checker correctness, and the worked example.  It currently contains about 9,400 lines of Lean across the main source files, including the JSON example.

The declared Lean toolchain is:
\begin{lstlisting}
leanprover/lean4:v4.29.0-rc6
\end{lstlisting}
The pinned mathlib revision is:
\begin{lstlisting}
5a2fd4957159dfe7444378d578b13a94f0122d8e
\end{lstlisting}
After cloning the repository at the pinned commit, the library and worked example are
checked with:
\begin{lstlisting}[language=bash]
lake build ConstrainedDecodingFormalization ConstrainedDecodingFormalization.GCDTest
\end{lstlisting}
The optional command \lean{lake exe cache get} downloads precompiled Mathlib files
and reduces build time, but is not required for correctness.
The development contains no \lean{sorry} and declares no axioms; \lean{native_decide} is used only to discharge the finite side conditions of the worked example (Section~\ref{sec:findings}). The public artifact is pinned to commit \lean{989fd77dc39b7d2e97bee8719a797efa3afac9fc}.

%% file: references.bib
@inproceedings{park2025flexible,
  author    = {Kanghee Park and Timothy Zhou and Loris D'Antoni},
  title     = {Flexible and Efficient Grammar-Constrained Decoding},
  booktitle = {Proceedings of the 42nd International Conference on Machine Learning},
  series    = {Proceedings of Machine Learning Research},
  volume    = {267},
  pages     = {48262--48275},
  publisher = {PMLR},
  year      = {2025},
  url       = {https://proceedings.mlr.press/v267/park25l.html}
}

@inproceedings{blaudeau2020packrat,
  author    = {Cl{\'e}ment Blaudeau and Natarajan Shankar},
  title     = {A Verified Packrat Parser Interpreter for Parsing Expression Grammars},
  booktitle = {Proceedings of the 9th ACM SIGPLAN International Conference on Certified Programs and Proofs},
  pages     = {3--17},
  publisher = {Association for Computing Machinery},
  year      = {2020},
  doi       = {10.1145/3372885.3373836}
}

@inproceedings{chattopadhyay2025lookaround,
  author    = {Agnishom Chattopadhyay and Angela W. Li and Konstantinos Mamouras},
  title     = {Verified and Efficient Matching of Regular Expressions with Lookaround},
  booktitle = {Proceedings of the 14th ACM SIGPLAN International Conference on Certified Programs and Proofs},
  pages     = {198--213},
  publisher = {Association for Computing Machinery},
  year      = {2025},
  doi       = {10.1145/3703595.3705884}
}

@inproceedings{jourdan2012lr1,
  author    = {Jacques-Henri Jourdan and Fran{\c{c}}ois Pottier and Xavier Leroy},
  title     = {Validating {LR(1)} Parsers},
  editor    = {Helmut Seidl},
  booktitle = {Programming Languages and Systems -- 21st European Symposium on Programming},
  series    = {Lecture Notes in Computer Science},
  volume    = {7211},
  pages     = {397--416},
  publisher = {Springer},
  year      = {2012},
  doi       = {10.1007/978-3-642-28869-2_20}
}

@inproceedings{rau2024earley,
  author    = {Martin Rau and Tobias Nipkow},
  title     = {A Verified Earley Parser},
  editor    = {Yves Bertot and Temur Kutsia and Michael Norrish},
  booktitle = {15th International Conference on Interactive Theorem Proving},
  series    = {Leibniz International Proceedings in Informatics (LIPIcs)},
  volume    = {309},
  pages     = {31:1--31:18},
  publisher = {Schloss Dagstuhl -- Leibniz-Zentrum f{\"u}r Informatik},
  address   = {Dagstuhl, Germany},
  year      = {2024},
  doi       = {10.4230/LIPIcs.ITP.2024.31}
}

@inproceedings{lasser2019ll1,
  author    = {Sam Lasser and Chris Casinghino and Kathleen Fisher and Cody Roux},
  title     = {A Verified {LL(1)} Parser Generator},
  editor    = {John Harrison and John O'Leary and Andrew Tolmach},
  booktitle = {10th International Conference on Interactive Theorem Proving},
  series    = {Leibniz International Proceedings in Informatics (LIPIcs)},
  volume    = {141},
  pages     = {24:1--24:18},
  publisher = {Schloss Dagstuhl -- Leibniz-Zentrum f{\"u}r Informatik},
  address   = {Dagstuhl, Germany},
  year      = {2019},
  doi       = {10.4230/LIPIcs.ITP.2019.24}
}

@inproceedings{lasser2021costar,
  author    = {Sam Lasser and Chris Casinghino and Kathleen Fisher and Cody Roux},
  title     = {{CoStar}: A Verified {ALL(*)} Parser},
  booktitle = {Proceedings of the 42nd ACM SIGPLAN International Conference on Programming Language Design and Implementation},
  pages     = {420--434},
  publisher = {Association for Computing Machinery},
  year      = {2021},
  doi       = {10.1145/3453483.3454053}
}

@inproceedings{edelmann2020zippy,
  author    = {Romain Edelmann and Jad Hamza and Viktor Kun{\v{c}}ak},
  title     = {Zippy {LL(1)} Parsing with Derivatives},
  booktitle = {Proceedings of the 41st ACM SIGPLAN Conference on Programming Language Design and Implementation},
  pages     = {1036--1051},
  publisher = {Association for Computing Machinery},
  year      = {2020},
  doi       = {10.1145/3385412.3385992}
}

@inproceedings{egolf2022verbatimpp,
  author    = {Derek Egolf and Sam Lasser and Kathleen Fisher},
  title     = {Verbatim++: Verified, Optimized, and Semantically Rich Lexing with Derivatives},
  booktitle = {Proceedings of the 11th ACM SIGPLAN International Conference on Certified Programs and Proofs},
  pages     = {27--39},
  publisher = {Association for Computing Machinery},
  year      = {2022},
  doi       = {10.1145/3497775.3503694}
}

@inproceedings{coquand2011regexp,
  author    = {Thierry Coquand and Vincent Siles},
  title     = {A Decision Procedure for Regular Expression Equivalence in Type Theory},
  editor    = {Jean-Pierre Jouannaud and Zhong Shao},
  booktitle = {Certified Programs and Proofs},
  series    = {Lecture Notes in Computer Science},
  volume    = {7086},
  pages     = {119--134},
  publisher = {Springer},
  year      = {2011},
  doi       = {10.1007/978-3-642-25379-9_11}
}

@inproceedings{geng2023gcd,
  author    = {Saibo Geng and Martin Josifoski and Maxime Peyrard and Robert West},
  title     = {Grammar-Constrained Decoding for Structured {NLP} Tasks without Finetuning},
  editor    = {Houda Bouamor and Juan Pino and Kalika Bali},
  booktitle = {Proceedings of the 2023 Conference on Empirical Methods in Natural Language Processing},
  pages     = {10932--10952},
  publisher = {Association for Computational Linguistics},
  address   = {Singapore},
  month     = {December},
  year      = {2023},
  doi       = {10.18653/v1/2023.emnlp-main.674},
  url       = {https://aclanthology.org/2023.emnlp-main.674/}
}

@inproceedings{poesia2022synchromesh,
  author    = {Gabriel Poesia and Alex Polozov and Vu Le and Ashish Tiwari and Gustavo Soares and Christopher Meek and Sumit Gulwani},
  title     = {Synchromesh: Reliable Code Generation from Pre-trained Language Models},
  booktitle = {International Conference on Learning Representations},
  year      = {2022},
  url       = {https://openreview.net/forum?id=KmtVD97J43e}
}

@misc{willard2023efficient,
  author        = {Brandon T. Willard and R{\'e}mi Louf},
  title         = {Efficient Guided Generation for Large Language Models},
  year          = {2023},
  eprint        = {2307.09702},
  archivePrefix = {arXiv},
  primaryClass  = {cs.CL},
  doi           = {10.48550/arXiv.2307.09702}
}

@inproceedings{beurerkellner2024domino,
  author    = {Luca Beurer-Kellner and Marc Fischer and Martin Vechev},
  title     = {Guiding {LLM}s The Right Way: Fast, Non-Invasive Constrained Generation},
  booktitle = {Proceedings of the 41st International Conference on Machine Learning},
  series    = {Proceedings of Machine Learning Research},
  volume    = {235},
  pages     = {3658--3673},
  publisher = {PMLR},
  year      = {2024},
  url       = {https://proceedings.mlr.press/v235/beurer-kellner24a.html}
}

@article{ugare2024syncode,
  author  = {Shubham Ugare and Tarun Suresh and Hangoo Kang and Sasa Misailovic and Gagandeep Singh},
  title   = {{SynCode}: {LLM} Generation with Grammar Augmentation},
  journal = {Transactions on Machine Learning Research},
  year    = {2025},
  issn    = {2835-8856},
  url     = {https://openreview.net/forum?id=HiUZtgAPoH}
}

@inproceedings{dong2024xgrammar,
  author    = {Yixin Dong and Charlie F. Ruan and Yaxing Cai and Ziyi Xu and Yilong Zhao and Ruihang Lai and Tianqi Chen},  
  title     = {{XGrammar}: Flexible and Efficient Structured Generation Engine for Large Language Models},
  booktitle = {Proceedings of Machine Learning and Systems},
  volume    = {7},
  publisher = {MLSys},
  year      = {2025},
  url       = {https://proceedings.mlsys.org/paper_files/paper/2025/hash/5c20ca4b0b20b0bd2f1d839dc605e70f-Abstract-Conference.html}
}

@inproceedings{demoura2021lean4,
  author    = {Leonardo de Moura and Sebastian Ullrich},
  title     = {The {Lean} 4 Theorem Prover and Programming Language},
  editor    = {Andr{\'e} Platzer and Geoff Sutcliffe},
  booktitle = {Automated Deduction -- {CADE} 28},
  series    = {Lecture Notes in Computer Science},
  volume    = {12699},
  pages     = {625--635},
  publisher = {Springer},
  year      = {2021},
  doi       = {10.1007/978-3-030-79876-5_37}
}

@inproceedings{mathlib2020,
  author    = {{The mathlib Community}},
  title     = {The {Lean} Mathematical Library},
  booktitle = {Proceedings of the 9th ACM SIGPLAN International Conference on Certified Programs and Proofs},
  series    = {{CPP} 2020},
  pages     = {367--381},
  publisher = {Association for Computing Machinery},
  address   = {New York, NY, USA},
  location  = {New Orleans, LA, USA},
  year      = {2020},
  isbn      = {9781450370974},
  doi       = {10.1145/3372885.3373824}
}

@inproceedings{li2026psc,
  author    = {Yongmin Li and Yihong Dong and Jia Li and Ge Li},
  title     = {Efficient Grammar-Constrained Decoding via Parser Stack Classification},
  booktitle = {Proceedings of the 35th ACM SIGSOFT International Symposium on Software Testing and Analysis},
  year      = {2026},
  note      = {To appear},
  eprint    = {2608.03065},
  archivePrefix = {arXiv},
  url       = {https://arxiv.org/abs/2608.03065}
}

@article{chen2021codex,
  author    = {Mark Chen and Jerry Tworek and Heewoo Jun and Qiming Yuan and Henrique Ponde de Oliveira Pinto and Jared Kaplan and Harri Edwards and Yuri Burda and Nicholas Joseph and Greg Brockman and Alex Ray and Raul Puri and Gretchen Krueger and Michael Petrov and Heidy Khlaaf and Girish Sastry and Pamela Mishkin and Brooke Chan and Scott Gray and Nick Ryder and Mikhail Pavlov and Alethea Power and Lukasz Kaiser and Mohammad Bavarian and Clemens Winter and Philippe Tillet and Felipe Petroski Such and Dave Cummings and Matthias Plappert and Fotios Chantzis and Elizabeth Barnes and Ariel Herbert-Voss and William Hebgen Guss and Alex Nichol and Alex Paino and Nikolas Tezak and Jie Tang and Igor Babuschkin and Suchir Balaji and Shantanu Jain and William Saunders and Christopher Hesse and Andrew N. Carr and Jan Leike and Josh Achiam and Vedant Misra and Evan Morikawa and Alec Radford and Matthew Knight and Miles Brundage and Mira Murati and Katie Mayer and Peter Welinder and Bob McGrew and Dario Amodei and Sam McCandlish and Ilya Sutskever and Wojciech Zaremba},
  title     = {Evaluating Large Language Models Trained on Code},
  journal   = {arXiv preprint arXiv:2107.03374},
  year      = {2021},
  url       = {https://arxiv.org/abs/2107.03374}
}

@inproceedings{schick2023toolformer,
  author    = {Timo Schick and Jane Dwivedi-Yu and Roberto Dess{\`\i} and Roberta Raileanu and Maria Lomeli and Eric Hambro and Luke Zettlemoyer and Nicola Cancedda and Thomas Scialom},
  title     = {Toolformer: Language Models Can Teach Themselves to Use Tools},
  booktitle = {Advances in Neural Information Processing Systems 36},
  year      = {2023},
  url       = {https://openreview.net/forum?id=Yacmpz84TH}
}

@inproceedings{patil2024gorilla,
  author    = {Shishir G. Patil and Tianjun Zhang and Xin Wang and Joseph E. Gonzalez},
  title     = {Gorilla: Large Language Model Connected with Massive {APIs}},
  booktitle = {Advances in Neural Information Processing Systems 37},
  year      = {2024},
  url       = {https://openreview.net/forum?id=tBRNC6YemY}
}

@article{gao2024texttosql,
  author    = {Dawei Gao and Haibin Wang and Yaliang Li and Xiuyu Sun and Yichen Qian and Bolin Ding and Jingren Zhou},
  title     = {Text-to-{SQL} Empowered by Large Language Models: {A} Benchmark Evaluation},
  journal   = {Proceedings of the VLDB Endowment},
  volume    = {17},
  number    = {5},
  pages     = {1132--1145},
  year      = {2024},
  doi       = {10.14778/3641204.3641221}
}

@inproceedings{first2023baldur,
  author    = {Emily First and Markus N. Rabe and Talia Ringer and Yuriy Brun},
  title     = {Baldur: Whole-Proof Generation and Repair with Large Language Models},
  booktitle = {Proceedings of the 31st ACM Joint European Software Engineering Conference and Symposium on the Foundations of Software Engineering},
  pages     = {1229--1241},
  publisher = {Association for Computing Machinery},
  year      = {2023},
  doi       = {10.1145/3611643.3616243}
}

@misc{bhat2026artifact,
  author       = {Manu Bhat and Arnav Dandu and Kanghee Park and Loris D'Antoni},
  title        = {Lean 4 Formalization of Machine-Checked Grammar-Constrained Decoding},
  year         = {2026},
  publisher    = {Zenodo},
  version      = {1.0},
  doi          = {10.5281/zenodo.24447388},
  url          = {https://doi.org/10.5281/zenodo.24447388},
  howpublished = {Software artifact, \url{https://github.com/large-loris-models/great-gramma-formalization}}
}
